\documentclass[10pt,journal]{IEEEtran}
\usepackage{amssymb}
\usepackage{graphicx}
\usepackage{epstopdf}
\usepackage{gensymb}
\usepackage{color}
\usepackage{amsmath}
\usepackage{bm}
\usepackage{etoolbox}
\usepackage{cite}
\usepackage{array}
\usepackage{booktabs}
\usepackage{multirow}
\usepackage{comment}
\usepackage{subcaption}
\usepackage[ruled,vlined]{algorithm2e}
\usepackage{algorithmicx}
\usepackage{mathrsfs}
\usepackage{algpseudocode}
\usepackage{hyperref}
\usepackage{balance}
\hypersetup{
    colorlinks=true,
    linkcolor=blue,
    filecolor=magenta,      
    urlcolor=cyan,
}

\usepackage[framemethod=TikZ]{mdframed}
 
\alglanguage{pseudocode}

\algnewcommand{\Initialize}[1]{
  \State \textbf{Initialize:}
  \Statex \hspace*{\algorithmicindent}\parbox[t]{.8\linewidth}{\raggedright #1}
}

\DeclareMathOperator*{\argmin}{\arg\!\min}
\usepackage{threeparttable}
\usepackage{mathtools}
\usepackage{hyperref}
\usepackage{pifont}
\usepackage{listings}

\DeclareMathOperator*{\argmax}{argmax}

\usepackage{url}
\urldef{\mailsa}\path|{alfred.hofmann, ursula.barth, ingrid.haas, frank.holzwarth,|
	\urldef{\mailsb}\path|anna.kramer, leonie.kunz, christine.reiss, nicole.sator,|
	\urldef{\mailsc}\path|erika.siebert-cole, peter.strasser, lncs}@springer.com|    

\usepackage[utf8]{inputenc}
\usepackage[english]{babel}

\usepackage{amsthm}

\usepackage{soul}

\theoremstyle{definition}

\definecolor{R}{RGB}{0,0,150}

\theoremstyle{remark}

\ifCLASSINFOpdf

\begin{document}
\title{RBNN: Memory-Efficient \underline{R}econfigurable Deep \underline{B}inary \underline{N}eural \underline{N}etwork with IP Protection for Internet of Things}

\author{
Huming Qiu\IEEEauthorrefmark{1}, Hua Ma\IEEEauthorrefmark{1}, Zhi Zhang, Yansong Gao\IEEEauthorrefmark{2}, Yifeng Zheng, Anmin Fu, \\ Pan Zhou, {\it Senior Member IEEE}, Derek Abbott, {\it Fellow IEEE}, and
Said F.~Al-Sarawi, {\it Member IEEE}.

\IEEEcompsocitemizethanks{\IEEEcompsocthanksitem H.~Qiu\IEEEauthorrefmark{1} and H. Ma\IEEEauthorrefmark{1} contributed equally.}

\IEEEcompsocitemizethanks{\IEEEcompsocthanksitem Y.~Gao\IEEEauthorrefmark{2} is the corresponding author.}

\IEEEcompsocitemizethanks{\IEEEcompsocthanksitem H.~Qiu, Y.~Gao and A.~Fu are with School of Computer Science and Engineering, Nanjing University of Science and Technology, Nanjing, China. e-mail: \{120106222682;yansong.gao;fuam\}@njust.edu.cn.}

\IEEEcompsocitemizethanks{\IEEEcompsocthanksitem  H.~Ma, S.~Al-Sarawi, and D.~Abbott are with School of Electrical and Electronic Engineering, The University of Adelaide, Adelaide, Australia. e-mail: \{hua.ma;said.alsarawi;derek.abbott\}@adelaide.edu.au}

\IEEEcompsocitemizethanks{\IEEEcompsocthanksitem Z.~Zhang is with Data61, CSIRO, Sydney, Australia. e-mail: zhi.zhang@data61.csiro.au.}

\IEEEcompsocitemizethanks{\IEEEcompsocthanksitem  Y.~Zheng is with the School of Computer Science and Technology, Harbin Institute of Technology, Shenzhen, Guangdong 518055, China. e-mail: yifeng.zheng@hit.edu.cn.}

\IEEEcompsocitemizethanks{\IEEEcompsocthanksitem  P.~Zhou is with the Hubei Engineering Research Center on Big Data Security, School of Cyber Science and Engineering, Huazhong University of Science and Technology, Wuhan, 430074, Hubei, China. e-mail: panzhou@hust.edu.cn.}

}
 
\maketitle

\begin{abstract}
Currently, a high demand for on-device deep neural network (DNN) model deployment is limited by the large model size, computing-intensive floating-point operations (FLOPS), and intellectual property (IP) infringements (i.e., easy access to model duplication for avoidance of license payments).
One appealing solution to addressing the first two concerns is model quantization, which reduces the model size and uses integer operations commonly supported by microcontrollers (MCUs usually do not support FLOPS). To this end, a 1 bit quantized DNN model or deep binary neural network (BNN) significantly improves the memory efficiency, where each parameter in a BNN model has only 1 bit. However, BNN cannot directly provide IP protection (in particular, the functionality of the model is locked unless there is a license payment).
In this paper, we propose a reconfigurable BNN (RBNN) to further amplify the memory efficiency for resource-constrained IoT devices while naturally protecting the model IP. Generally, RBNN can be reconfigured on demand to achieve any one of $M$ ($M>1$) distinct tasks with the same parameter set, thus only a single task determines the memory requirements. 
In other words, the memory utilization is improved by a factor of $M$. 
Our extensive experiments corroborate that up to seven commonly used tasks ($M=7$, six of these tasks are image related and the last one is audio) can co-exist (the value of $M$ can be larger). These tasks with a varying number of classes have no or negligible accuracy drop-off (i.e., within 1\%) on three binarized popular DNN architectures including VGG, ResNet, and ReActNet. The tasks span across different domains, e.g., computer vision and audio domains validated herein, with the prerequisite that the model architecture can serve those cross-domain tasks. To fulfill the IP protection of an RBNN model, the reconfiguration can be controlled by both a user key and a device-unique root key generated by the intrinsic hardware fingerprint (e.g., SRAM memory power-up pattern). By doing so, an RBNN model can only be used per paid user per authorized device, thus benefiting both the user and the model provider. The source code is released at \textcolor{blue}{\url{https://github.com/LearningMaker/RBNN}}.

\end{abstract}

\IEEEpeerreviewmaketitle



\section{Introduction}\label{sec:Intro}
Deep neural networks (DNNs) today have been deployed in numerous tasks such as computer vision, natural language processing, and speech recognition. With the trend of Machine Learning as a Service (MLaaS), trained models are put onto the cloud to provide inference services. However, many critical tasks (e.g., autonomous driving) require zero- or minimum-latency in the prediction or inference stage, thus the latency caused by the interaction between the cloud and user is a hurdle.
In addition, when a model is hosted on the cloud, it is limited to a fixed level of network bandwidth that might not always be achievable. 
Further, from the privacy perspective, hosting the model on the cloud poses a threat to data privacy, e.g., data must not leave the device where the user does not want to send the private data to the server. To achieve low-latency, reliable network connectivity, and privacy\footnote{Please note that privacy leakage can occur in various vectors in deep learning. Performing model inference on local devices can address the privacy concern where user data are sent to the cloud for inference, while other privacy concerns~\cite{boulemtafes2020review} may still exist, e.g., knowledge distillation-based privacy leakage~\cite{kundu2021analyzing}.}, a straightforward solution is to host the model on localized pervasive mobile computing platforms such as internet of things (IoT) devices. 

However, a DNN model has a large number of parameters that are typically stored with a precision of \texttt{float32}, demanding a large model memory footprint. For instance, AlexNet~\cite{krizhevsky2017imagenet} has more than one hundred million parameters that can potentially result in a model size of 240\,MB~\cite{iandola2016squeezenet}. Such memory requirements can hardly be satisfied by memory-constrained devices. 
We note that FPGA is commonly utilized today for accelerating the DNN inference while its on-chip memory can be as low as 10\,MB~\cite{iandola2016squeezenet}.
Furthermore, floating-point operations (FLOPS) are numerous and computing-intensive. Take for example the ResNet-50~\cite{he2016deep}, which needs about 4.1B FLOPS to process an image of size 224$\times$224$\times$3~\cite{han2020ghostnet}. The FLOPS are available in high-performance hardware such as GPUs but may not be (well) supported by low-power IoT devices such as drones and watches.  

The above requirements of large model size and extensive FLOPS lead to heavyweight DNN models as well as  notable model inference/prediction latency, thus hindering the DNN model's ubiquitous deployment on IoT devices. As such, the model must be as lightweight as possible~\cite{zhang2018shufflenet} when it is implemented in resource-constrained computing platforms. A practical solution is to reduce model precision by performing so-called \textit{model quantization}. This approach has attracted significant attention from both academia~\cite{david2020tensorflow} and industry~\cite{TensorflowLite}.

\subsection{Limitations:} 

\noindent\textbf{Model Memory Efficiency.} Model quantization can be classified into two categories based on the phase in which quantization is introduced: post-training quantization and quantization-aware training~\cite{TensorflowLite}. The former quantizes the weight and/or activation after the model has been trained and results in a slight degradation in the inference accuracy, which is usually applicable when quantizing heavyweight DNN models. If a model is designed to be lightweight to cater for IoT devices, then quantization-aware training improves accuracy, compared to post-training quantization.
The quantization of these models can range from 8-bit, 4-bit and even down to 1 bit~\cite{hubara2016binarized,rastegari2016xnor,bulat2019xnor,martinez2020training,andri2017yodann,conti2018xnor} for not only weight but also activation~\cite{qin2020binary}. 
Particularly, 1-bit quantized models are referred to as binary neural networks (BNNs) where both the weights and activations are represented by two possible values, -1/0 and +1, substantially reducing the memory footprints. In addition, the FLOPS are replaced by simpler operations such as the XNOR logical operation and bitcount to expedite the inference~\cite{zhang2020fracbnn}. 

To this end, \textit{it seems that the memory efficiency for a given model is maximized by the binarized model}. We are now interested in the following research question:

\begin{quote}
Is it possible to further increase the memory utilization of a given BNN? If so, what evidence can be used to pragmatically demonstrate that?
\end{quote}

\noindent\textbf{Model IP Protection.} At the same time, on-device model (in particular, BNN in this work) deployment adversely raises a critical security concern. In particular, the unauthorized user may replicate the same model---due to easy access to the model---to other devices or even share the model with many others without proprietary payments, imposing monetary loss to the model provider. This is because the model training could involve incredible computational resources, large proprietary data, and domain expertise, thus possessing significant commercial value to the model provider~\cite{adi2018turning,darvish2019deepsigns,lin2020chaotic,chakraborty2020hardware}. Therefore, we are further interested in the following research question:

\begin{quote}
Is it possible to protect the on-device model IP by preventing unauthorized usage, while increasing the model memory utilization?
\end{quote}

\subsection{Our Solution} We propose a reconfigurable deep binary neural network (RBNN) to integrate various tasks into the same model, thus further improving the memory utilization to facilitate on-device model deployment, and protecting the model IP at the same time. 
Our main insight is that multiple tasks can be concurrently learned during the training stage through proper objective functions. Thus, multiple tasks can be trained at the same time on the same model~\cite{ruder2017overview}. However, unlike common multi-task learning that shares representations among related tasks~\cite{ruder2017overview}, the RBNN does \textit{not} share representations and thus \textit{can} be trained for completely different tasks. By {using different model parameter permutations, the RBNN can change the inference task on-demand to serve multiple purposes}.
Take the smart home as an example, we can train a face recognition task and a speech recognition task in the same RBNN model and both tasks share the same model parameters including weights and activations. At the inference stage, the trained model first serves face recognition---multiple trials can be allowed if inference accuracy is low---to authorize the user to enter the home and then switch to the speech recognition task to serve as a voice assistant.
At the same time, we recognize that the reconfigurable nature of the RBNN is a convenient feature for preventing the model being used by unauthorized users or/and even devices, as the reconfiguration can be bound up with an authorized key from the model provider to enable IP protection on a deployed RBNN model, that is, only a unique key can properly configure/unlock the model to a specific paid task.

\subsection{Contributions} 
Our experiments on RBNN are to answer the following research questions.

\begin{mdframed}[backgroundcolor=black!10,rightline=false,leftline=false,topline=false,bottomline=false,roundcorner=2mm]
    Q1: Can multiple distinctive tasks (e.g., with varying number of classes/labels or/and from different domains) be efficiently co-learned during the training, thus enabling RBNN?
\end{mdframed}

\begin{mdframed}[backgroundcolor=black!10,rightline=false,leftline=false,topline=false,bottomline=false,roundcorner=2mm]
    Q2: Can the IP protection functionality be enabled or integrated with RBNN?
\end{mdframed}

Building upon the experiments designed for above questions, contributions of this work are threefold:
\begin{enumerate}
    \item We propose RBNN to address memory constraints when deploying a DNN model onto IoT devices to achieve low-latency and low-network connectivity. Instead of training a BNN model for a single task, {a single RBNN model can be reconfigured on-demand to serve multiple inference tasks}.
    \item We comprehensively evaluate RBNN on three common model architectures including VGG, ResNet, and ReActNet (with ResNet and MobileNet as backbone respectively) on seven commonly used datasets. We demonstrate that i) tasks with a varying number of classes can be integrated into the RBNN, ii) tasks from two different domains: vision and audio, can co-exist in the RBNN as the model architecture can handle cross-domain tasks. In addition, RBNN is independent on model architectures as validated via three varying model architectures. Our experiments show that RBNN achieves superior memory efficiency and the accuracy for each inference task in the RBNN is on par with its corresponding BNN ({\bf Q1}).
    \item We propose an IP protection framework to bind the RBNN reconfiguration with both \textit{a user key} and \textit{a device key} derived from a physical unclonable function (PUF) to prevent executing the model on an unauthorized device, and enabling the IP license to be issued per device and per user ({\bf Q2}).
\end{enumerate}

The remainder of this paper is organized as follows. Section~\ref{sec:relatedwork} gives the necessary background related to the presented work. Section~\ref{sec:reconRBNN} details the realization of the proposed RBNN framework. Section~\ref{sec:experiment} comprehensively evaluates the RBNN with various model architectures and datasets. Section~\ref{sec:IPProtection} exploits the reconfiguration to naturally protect the RBNN IP by binding the reconfiguration of each task by a unique key. Section~\ref{sec:CompareAndDiscuss} compares the RBNN with other related designs from the multi-task integration and security aspects, respectively, and provides further discussions.
This work is concluded in Section~\ref{sec:Conclusion}.

\section{Background and Related Work}\label{sec:relatedwork}
\subsection{Deep Neural Network}\label{Sec:DNNDef}
When provided with an input $x\in \mathbb{R}^n$ with $n$ dimensions, a DNN $F_{\theta}$ will map $x$ to one of $C$ classes in common classification tasks. More precisely, the output $y\in \mathbb{R}^m$ of the DNN is a so-called softmax---a probability vector for $C$ classes.
In other words, $y_i$ is the probability that the input $x$ belongs to the $i_{\rm th}$ class. An input $x$ is inferred as class $i$ if $y_i$ has the highest probability, and thus the output class label $z$ is $\argmax_{i \in [1,C]} y_i$~\cite{gao2019strip}. 

Generally, the DNN model $F_{\theta}$ is trained with a number of samples with annotated ground-truth labels---this is the training dataset. The trainable parameter ${\theta}$ comprises weights and biases, which will be learned during the training phase. Specifically, suppose that the training dataset is a set, $\mathcal{D}_{\rm train} = \{x_i, y_i\}_{i=1}^{S}$ of $S$ samples, where $x_i \in \mathbb{R}^n$ and corresponding ground-truth labels $z_i \in [1, C]$. 
The training phase permits learning of parameters that minimize the errors between inputs' predictions and their ground-truth labels, which can be evaluated by a loss function $\mathcal{L}$. Parameters $\Theta$ after training are expressed as:
\begin{equation}\label{Eq:parameter}
    \Theta = \argmin_{\Theta^*} \sum_i^S \mathcal{L}(F_{\Theta}(x_i), z_i). 
\end{equation}

Notably, to have accurate inference, the DNN always uses many layers. Thus, a very large number of parameters need to be trained and stored. In the inference stage, the computation and storage overhead are highly related to the size of the DNN model. In addition, the parameters are expressed with \texttt{float32} precision by default, so inference is performed based on the intensive floating-point operations (FLOPS). 

\begin{figure*}[t]
	\centering
	\includegraphics[trim=0 0 0 0,clip,width=0.70\textwidth]{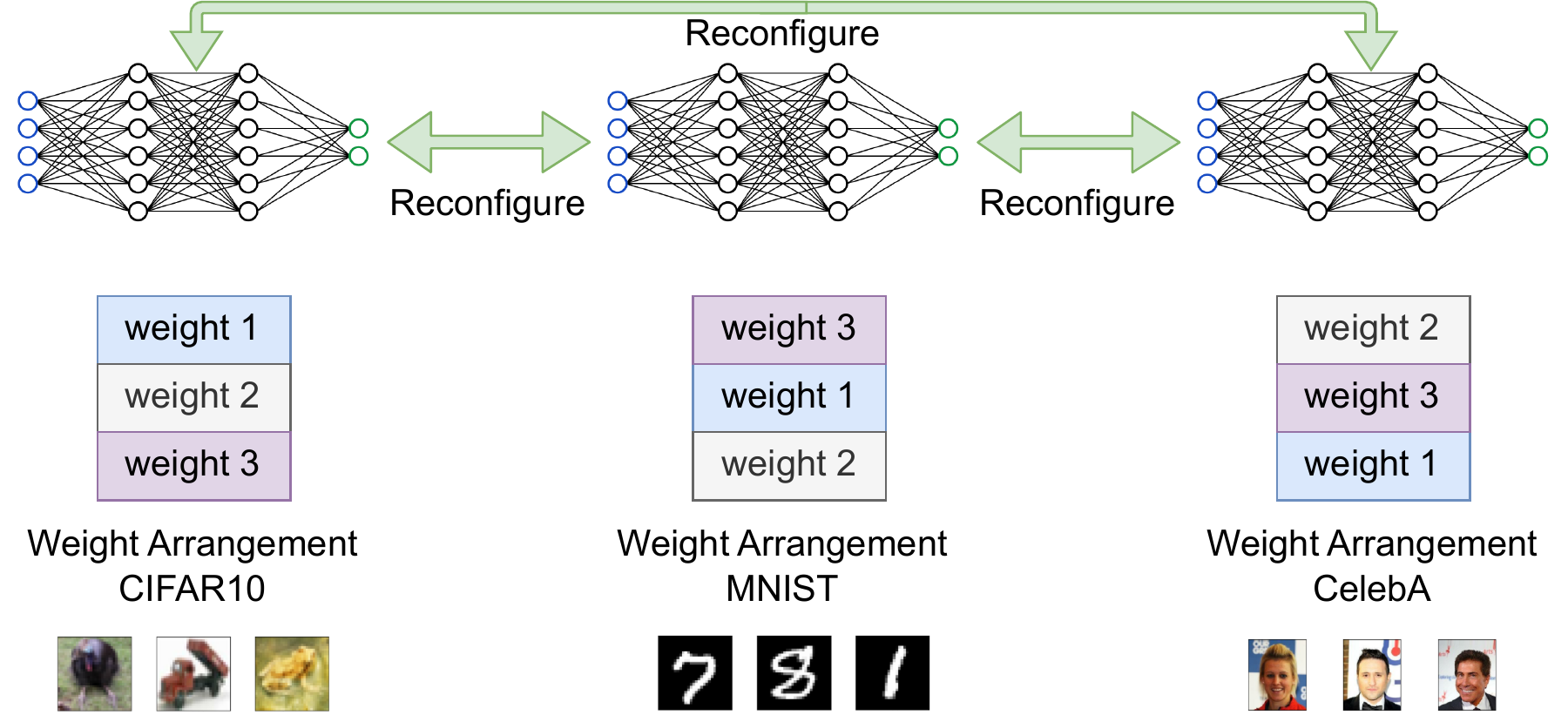}
	\caption{Overview of reconfigurable deep (binary) neural network. Rearranging the same network parameters (i.e., weight and activation) can switch to a network task from e.g., natural image classification on CIFAR10 to digit recognition on MNIST. Notably, the parameter rearrangement can be individually applied to \textit{each layer}. The bidirectional arrow means that the \textit{any two tasks} can be flexibly reconfigured from one to the other.}
	\label{fig:overview}
\end{figure*}

\subsection{Binary Neural Network}
In the literature, a number of model compression techniques have been proposed to reduce the memory footprint and the computational overhead. These techniques include pruning or trimming off less important parameters~\cite{han2015deep,kundu2021dnr}, knowledge distillation for transferring knowledge from a large model to a small one~\cite{hinton2015distilling}, and model quantization~\cite{polino2018model}. This work focuses on the last item. 
Compared with a full-precision model that uses floating points, a quantized model compresses the original model weights by representing the model weights with lower precision while retaining the inference accuracy. The quantized model can utilize homogeneous-precision or mixed-precision quantization~\cite{kundu2022bmpq}. The ultimate quantization is binarization, resulting in BNN model, which plays a significant role in maximizing memory utilization for resource-constrained IoT devices.

Binarization is a 1-bit quantization, where a parameter has only two possible values, namely -1/0 or +1. When training a BNN model,
only 1 bit is used to represent both weight $w$ and activation $a$ parameters. Following~\cite{qin2020binary}, binary functions are expressed as follows:

\[Q_w(w)=\alpha b_w,\ Q_a(a)=\beta b_a,\]

\noindent where \(b_w\) and  \(b_a\) represent the binary weight and activation, and each is multiplied by a scale factor of \(\alpha\) or \(\beta\). The sign functions of \(Q_w\) and \(Q_a\) for binarization are as follows:

\begin{equation}\label{eq:sign}
\textsf{sign}(x)= \begin{cases} 
      +1 & x\geq 0 \\
      -1 & \text{otherwise}. 
      \end{cases}
\end{equation}

After binarization, the floating-point convolution operation in forward propagation can be expressed as:

\[z=\sigma(Q_w(w) \otimes Q_a(a))=\sigma(\alpha\beta(b_w \odot b_a)), \] 

\noindent where \(\odot\) represents the vector inner product realized by bitwise \textsf{XNOR} and \textsf{Bitcount} operations, which can be performed more efficiently. Thus, binarization not only saves memory footprint but also makes computation much more efficient as floating point operations are replaced by bitwise operations.

\subsection{Model IP Protection}

The IP protection of a DNN model is demanded in practice~\cite{adi2018turning,darvish2019deepsigns,lin2020chaotic,chakraborty2020hardware}. This is mainly because DNN models are usually trained with computing-intensive resources to process
a large amount of proprietary data. In addition, the model training also needs domain expertise. The trained model is valuable for a model provider as an IP and
should be protected to preserve its competitive commercial advantage~\cite{darvish2019deepsigns}.

Watermarking is an approach to embedding provider-known-secret information into the released model to prevent model piracy~\cite{adi2018turning,darvish2019deepsigns}, since later the model provider can use the watermark to claim the ownership. However, watermarking cannot prevent covertly leaked model usage, especially on the localized IoT devices. This trivially bypasses the ownership inspection through the technique of watermarking.

An alternative is to leverage obfuscation and locking to further enhance the security of model IP~\cite{xu2018deepobfuscation,chakraborty2020hardware}. Xu \textit{et al.}~\cite{xu2018deepobfuscation} structurally obfuscate the \textit{network architecture} to prevent \textit{architecture steal}.  Instead of obfuscating the model architecture, Chakraborty \textit{et al.}~\cite{chakraborty2020hardware} obfuscate parameters of the model, which can only exhibit high inference performance unless an end-user has legitimate access to unlock it. Without knowledge of the key being fed into the DNN, any unauthorized usage of the locked model shows substantial accuracy drop, which is useless for the attacker.

It is recognized that our RBNN naturally enables the model locking/obfuscating property by associating the reconfiguration with a key, thus providing an very useful security benefit to protect the model IP.

\subsection{Backdoor Attack by Abusing Multi-task Learning}
A backdoor attack introduces malicious behavior or tasks besides the main task for which the model is trained~\cite{gao2019strip,gao2020backdoor}. In other words, a backdoored model essentially has multiple tasks: a main task and one or more backdoored tasks~\cite{gao2020backdoor,bagdasaryan2020blind,guo2020trojannet,bagdasaryan2020backdoor}. The model trained for two tasks will exhibit both normal and backdoor behaviors, where the backdoor behavior is activated by the so-called \textit{trigger}---a secret known only to the adversary. 
While previous backdoored tasks share the representation with the main task,
TrojanNet~\cite{guo2020trojannet} is a new backdoor technique for full-precision models rather than BNN that abuses the multi-task learning property to make each backdoored task independent from the main task.

As summarized by Gao \textit{et al.}~\cite{gao2020backdoor}, the backdoor can also be used for legitimate purposes such as watermarking for model IP protection~\cite{adi2018turning}, acting as a honey-pot for trapping and catching adversarial examples~\cite{shan2020gotta}, and verification for data deletion requested by users who have contributed the data~\cite{sommer2020towards}. 
Our proposed RBNN employs the inherent multi-task property abused by the backdoor attack for legitimate use for resource-constrained IoT devices, e.g., improving memory efficiency of BNN-based models by embedding multiple tasks into a BNN model.

\section{Reconfigurable Deep BNN}\label{sec:reconRBNN}
\subsection{Overview}

Fig.~\ref{fig:overview} presents an overview of our proposed reconfigurable deep binary neural network. By rearranging the order of network parameters, the same network with the same parameters executes completely different tasks. This requires that multi-tasks be trained upon distinct datasets: each dataset corresponding to a specific task. It is noted that the multi-task training in RBNN does not learn shared representation among similar tasks as previously reported~\cite{ruder2017overview}, rather, it learns different representations specific to each task. Nonetheless, both cases leverage the high capacity or redundancy of the DNN. Below we elaborate on how to realize RBNN through binarization. 

\subsection{Realization}
\subsubsection{Multi-Task Training}
The key insight of RBNN is to utilize the multi-task learning capability of DNN. Thus, different tasks $\{T_1, T_2, \cdots, T_M\}$ share the same parameter set including weights and biases. For convenience, we use $T_m$ ($m\in\{1,\cdots,M\}$) to denote a network task. As illustrated in Fig.~\ref{fig:overview}, each $T_m$ can be immediately switched on by reconfiguring the network parameter order (i.e., parameter permutation).

Suppose that a dataset $\mathcal{D}_m$ is used to train $T_m$ and $(x_i^m, y_i^m)$ denotes the $i$-th sample from $\mathcal{D}_m$. The overall loss $L$ (see loss description in Eq.~\ref{Eq:parameter}) of the multi-task training is the summation of individual task losses, which can be expressed as below:

\begin{equation}\label{eq:loss}
\begin{split}
L & = \sum_{m=1}^{M} \underbrace{L_{T_m} (T_m({x}_i^m), y_i^m}_{L_{T_m}})\\ &  = \underbrace{ L_{T_1} (T_1(x_i^1), y_i^1)}_{L_{T_1}} + \cdots +
\underbrace{L_{T_M} (T_M({x}_i^M), y_i^M}_{L_{T_M}}).
\end{split}
\end{equation}

The overall loss $L$ can be optimized with gradient descent on ${\bf w}_{T_m}$ and its gradient is given by:
\begin{equation}
    \frac{\partial L}{\partial w} = \sum_{m=1}^{M} \frac{\partial L_{T_m}}{\partial {\bf w}_{T_m}}.
\end{equation}

\noindent The crucial step is to reconfigure the ${\bf w}$ to fit  $T_m$ and obtain $L_{T_M}$. This can be achieved by permutation, for example ${\bf w}_{T_2}\leftarrow \textsf{Reconfig}({\bf w}_{T_1})$ by reconfiguring the same parameters through permutation. In principle, we can train an arbitrary number of distinct tasks by the parameter permutations. In Eq.~\ref{eq:loss}, the importance of each task is equal, thus the regularization is same. 
Alternatively, we can use a distinct regularization factor for each $L_{T_m}$ to reflect the distinct importance of ${T_m}$ if needed.

Intuitively, we can store the permutation or the corresponding mask that corresponds to a task to facilitate the reconfiguration of \textsf{Reconfig}. However, this could introduce a non-negligible memory overhead. To address this issue, we customize the approach from~\cite{guo2020trojannet} and generate the \textsf{Reconfig} in the RBNN. Specifically, the tasks are reconfigured through distinct keys: each task is bound up with a specific key (i.e., $k_m$). Here, $k_m$ is used as the initial seed fed into a recursive function, preferably a pseudo-random number generator (PRNG) \textsf{H} albeit not a must, to produce recursive seeds that determines the permutation per subsequently network layer. Leveraging a recursive function, specifically \textsf{H}, could ensure the recursive seed determined permutation per network layer is preferably distinct, thus a different reconfiguration even per layer. We have validated that our RBNN implementation is insensitive to the specific recursive function implementation, regardless of PRNG based or not, though the former is preferable.

In this way, \textsf{Reconfig} is associated with $k_m$. As such, ${\bf w}_{T_m}\leftarrow \textsf{Reconfig}({\bf w}, k_m)$. A byproduct is that ${\bf w}_{T_m}$ can only be configured correctly with a valid $k_m$. If an invalid key is applied, the parameters in the network will be wrongly permuted, causing the neural network to malfunction. To this end, the reconfiguration with valid keys issued from the model provider can provide IP protection to the network (see details in Section~\ref{sec:IPProtection}).

As in Eq.~\ref{eq:loss}, reconfigurable training requires training multiple different datasets in the same neural network. After training one batch of one dataset, the RBNN trains a data batch from the next task's dataset in the \textit{same epoch}. 
More specifically, the RBNN interleaves data from multiple tasks during learning to jointly optimize the model weights as a way to avoid catastrophic forgetting that could occur when \textit{a following task is trained after the previous task is completely trained}~\cite{kirkpatrick2017overcoming}. 
When training {alternates with dataset batches}, $k_m$ is reconfigured to feed into \textsf{H} and transform ${\bf w}$ to ${\bf w}_{T_m}$ so that ${\bf w}_{T_m}$ corresponds to $\mathcal{D}_m$ for $T_m$.

\subsubsection{RBNN Training}
When training an RBNN model, the backpropagation algorithm based on gradient descent can be used directly to update the parameters~\cite{qin2020binary,hubara2016binarized}. However, the binarization function such as the \textsf{sign} function in Eq.~\ref{eq:sign} is usually not differentiable. Even worse, the derivative value of part of the function disappears, and the derivative of the function is almost $0$ everywhere, which is obviously incompatible with backward propagation.

Thus, the original gradient descent based backward propagation cannot be immediately used for updating the binarized weights. We solve this issue by using a pass-through estimator (STE) in backward propagation~\cite{hinton2012neural}. The derivative of STE can also be represented as the propagation gradient through the hard \textsf{tanh}, which is defined as:
\begin{equation}\label{eq:htanh}
    \textsf{Htanh}(x)={\textsf{clip}}(x,-1,1)=\textsf{max}(-1,\textsf{min}(1,x)).
\end{equation}

By applying the hard \textsf{tanh}, \textsf{Htanh}, activation function, the RBNN model can now be directly trained using the gradient descent that is same when training the full-precision models. Nonetheless, 
if the absolute value of the full-precision activation is larger than 1, the \textsf{Htanh}(x) cannot be updated in backward propagation. Therefore, the identity function (i.e. \(f(x)=x\)) is also chosen to approximate the derivative of the \textsf{sign} function in practice.

\section{Experiments}\label{sec:experiment}
The experiments in this section are designed mainly to answer the question ({\bf Q1}) listed in Section~\ref{sec:Intro}. We consider seven common datasets corresponding to six computer vision domain tasks and one audio domain task. The number of classes for each task ranges from 2 to 100. We use three common but different model architectures for a comprehensive evaluation.

\vspace{0.2cm}

\subsection{Setup}

\begin{table}
	\centering 
	\caption{Datasets Summery}
			\resizebox{0.4\textwidth}{!}{
	\begin{tabular}{c| c | c | c } %
		\toprule 
				
		Dataset &  \begin{tabular}{@{}c@{}} $\#$ of  \\ labels \end{tabular}  & \begin{tabular}{@{}c@{}} Image  \\ size\end{tabular} & \begin{tabular}{@{}c@{}} $\#$ of samples (train; test) \end{tabular} \\ 
		\midrule
		MNIST~\cite{lecun1998gradient} &  10 & $28\times28\times 1$ &  70,000 (60,000; 10,000) \\ \hline
        Fashion-MNIST~\cite{xiao2017fashion} &  10 & $28\times28\times 1$ &  70,000 (60,000; 10,000) \\ \hline
        SVHN~\cite{netzer2011reading} &  10 & $32\times32\times 1$ &  99,289 (73,257; 26,032) \\ \hline
        CIFAR10~\cite{krizhevsky2009learning} &  10 & $32\times32\times 3$ &  60,000 (50,000; 10,000) \\ \hline
        CIFAR100~\cite{krizhevsky2009learning} &  100 & $32\times32\times 3$ &  60,000 (50,000; 10,000) \\ \hline
        CelebA~\cite{liu2015faceattributes} &  2 & $218\times178\times 3$ &  182,732 (162,770; 19,962) \\ \hline
        \begin{tabular}{@{}c@{}}  Speech \\ Command (SC) \end{tabular}~\cite{speechcommandsv2} &  35 & $32\times32\times 1$ &  95,394 (57,294; 38,100) \\ 
		\bottomrule
	\end{tabular}
			}
	\label{tab:dataset}
\end{table}

\subsubsection{Datasets}
There are six image datasets under the vision domain and one speech recognition dataset under the audio domain, as summarized in Table~\ref{tab:dataset}. We then demonstrate that those different domain tasks can be reconfigured as long as the same model architecture can handle these tasks.

\vspace{0.15cm}
\noindent\textbf{MNIST/Fashion-MNIST:} The MNIST has handwritten digits of 10 classes by different people~\cite{lecun1998gradient}. The Fashion-MNIST dataset~\cite{xiao2017fashion} is composed of 10 types of fashion products. Both contain 60,000 samples for training and 10,000 samples for testing. Each sample image is a gray-scale or single channel image with the size of $28\times28$ pixels.

\vspace{0.15cm}
\noindent\textbf{SVHN:} This dataset is composed of street view house numbers from Google~\cite{netzer2011reading}. The training set contains 73,257 samples and the test set contains 26,032 samples.

Each gray-scale image has a size of $32\times 32$. Similar to MNIST, the SVHN is also a digital recognition dataset, but more difficult to recognize because of natural scenes.

\vspace{0.15cm}
\noindent\textbf{CIFAR10/100:} Both CIFAR10 and CIFA100 are natural color image datasets, with CIFAR10 and CIFAR100 having 10 and 100 classes, respectively~\cite{krizhevsky2009learning}. The training set and the testing set for both CIFAR10 and CIFAR100 contain 50,000 and 10,000 images respectively. 
The CIFAR100 is harder to recognize than CIFAR10 due to the increased number of classes and decreased number of training samples per class.

\vspace{0.15cm}
\noindent\textbf{CelebA:} This dataset contains large-scale face attributes with more than 200K celebrity images, each with 40 attribute annotations\cite{liu2015faceattributes}. The images in this dataset cover large pose variations and background clutter. CelebA has large diversities, large quantities, and rich annotations, including 10,177 identities, 202,599 face images, and 5 landmark locations, 40 binary attributes annotations per image. In our work, the training set and the testing set are divided according to the {list\_eval\_partition.txt} file in the dataset---we do not use the validation set in this file. The training and testing sets have 162,770 and 19,962 samples, respectively. We have selected the gender attribute as a {binary classification problem} to train the model.

\vspace{0.15cm}
\noindent\textbf{Speech Command (SC):} This audio dataset has spoken words designed to help train and evaluate keyword spotting systems~\cite{speechcommandsv2}. It consists of 105,829 utterances of 35 words. Each utterance is stored as a one-second (or less) WAVE format file, with the sample data encoded as linear 16-bit single-channel PCM (Pulse-code modulation) values, at a 16~kHz rate. Note that WAVE format files with a duration of less than one second are not used by us---these samples are excluded. As a consequence, a total of 95,394 audio data samples are left and randomly selected. 
Note that 60\% of the data is used as the training set, and the rest is the testing set. Before feeding into the model, the Mel-Frequency Cepstral Coefficients (MFCC) is used to extract speech features and convert them into a frequency spectrum. The sampling rate is set to 6,300~Hz in order to convert the audio data format to $32\times 32 \times 1$, and the mono data is copied to convert it into three channels. 

\subsubsection{Model Architecture}\label{sec:modelArchitecture}
We consider binarizing three types of DNNs: ResNet18~\cite{he2016deep}, VGG~\cite{simonyan2014very} and ReActNet (using both ResNet14 and MobileNet as backbone, respectively)~\cite{liu2020reactnet} for experimental validation. Note that ReActNet is a newly devised BNN framework to boost the BNN inference accuracy\footnote{The source code at \url{https://github.com/liuzechun/ReActNet} is adopted}. It keeps the fully connected layers with full-precision representation by design, which we have followed. In addition, the ResNet backbone of ReActNet in our work is ResNet14, where the last block of ResNet18 is removed. This ResNet14 setting along with MobileNet backbone use are to test whether the smaller model's redundancy still supports the reconfiguration without a drop in accuracy when multiple tasks are simultaneously learned.

For the VGG model, the architecture is \textsf{conv}(128)→\\\textsf{conv}(128)→\textsf{conv}(256)→\textsf{conv}(256)→\textsf{conv}(512)→\textsf{conv}(512)
→ \textsf{fc} (1024) → \textsf{fc}(1024) → \textsf{fc} (number of classes) where \textsf{conv}($\cdot$) is a convolutional layer with $3\times 3$ kernel size, and \textsf{fc}($\cdot$) is a fully connected layer. For the ResNet model, the specific ResNet18 is used. For both VGG and ResNet models, all convolutional layers and fully connected layers are binarized, and a batch normalization layer is added before the activation layer and after the fully connected layer. For the ReActNet model, the ResNet14 and MobileNet are used as the basic skeleton of the network. The ReActNet by design reserves fully-connected layer to be represented with full-precision weights instead of being binarized. 
For all three models, the ADAM optimizer is used. The batch size is 256. We train all tasks with 100 epochs. The initial learning rate is $5\times 10^{-3}$, which decreases to $1\times 10^{-3}$ at the $40_{\rm th}$ epoch and $5\times 10^{-4}$ at the $80_{\rm th}$ epoch.

We sequentially integrate six tasks under image domain (MNIST, Fashion-MNIST, SVHN, CIFAR10, CelebA, CIFAR100) to train each of the aforementioned model architectures to verify the RBNN performance (note that the RBNN is independent of the task adding sequence as detailed in Section~\ref{sec:taskadding}). In addition, we have added one more audio task, SC, when training the RBNN with ReActNet. This is to demonstrate the practicality of co-learning tasks from different domains as long as the specific model is capable of handling those tasks. The switch of each seven tasks is controlled with a distinctive key. In this way, the specific task can be configured immediately by using its key during the inference phase. 

\subsubsection{Input Size Alignment}
The image size of different classification tasks differs, see Table~\ref{tab:dataset}. For all the model architectures studied in this work, the model accepted image size is $32\times 32$. In this context, for image size that is unequal to $32\times 32$, we resize it to the model accepted input size. In addition, the model input image uses three channels by default. Therefore, for images, e.g., MNIST, SVHN as well as the spectrum of SC, with a single channel, we have firstly transformed them into three channels\footnote{The \texttt{torchvision.transforms.Grayscale(3)} command is used. In our case, this means that the single-channel image is replicated into a three-channel image.} before feeding them into the models.
\subsection{Experimental Results}

\begin{figure}[t]
	\centering
	\includegraphics[trim=0 0 0 0,clip,width=0.50\textwidth]{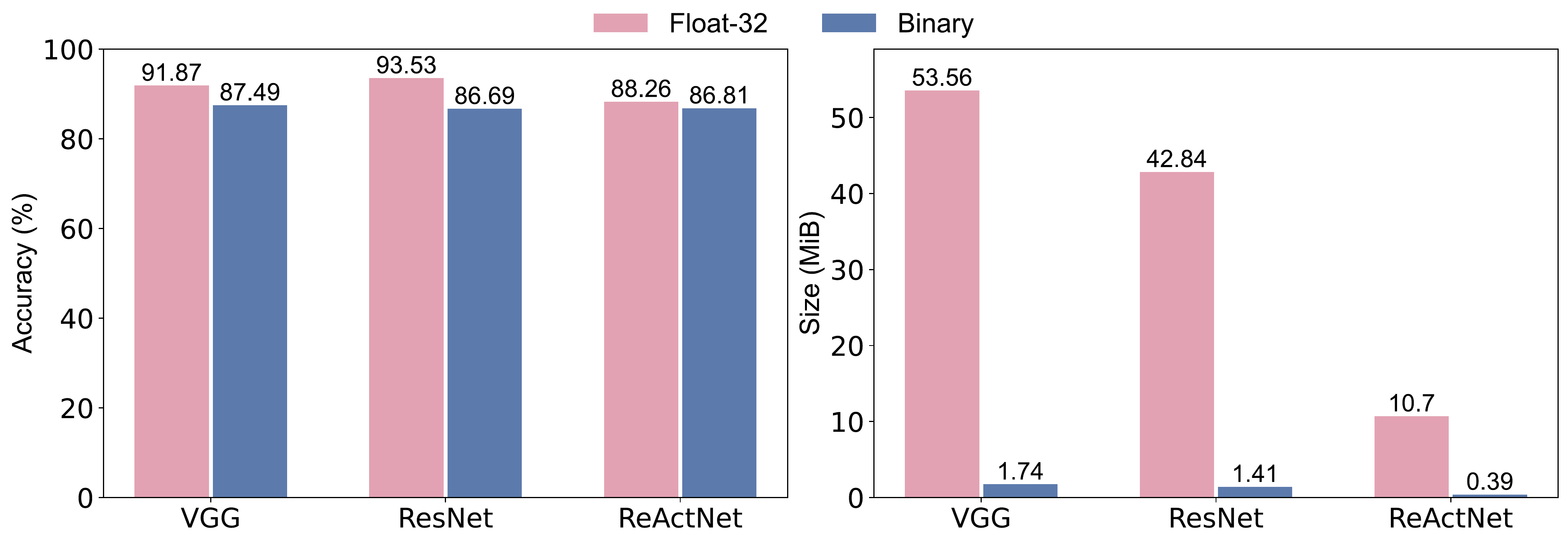}
	\caption{(Left) Testing accuracy of full precision model and BNN model of the same architectures. (Right) Model size of full precision model and BNN model of the same architectures.}
	\label{fig:Baseline}
\end{figure}

\subsubsection{Baseline} \noindent{\it Baseline.} To set a baseline, we represent the model parameters with full precision (\texttt{float32}) and 1-bit, respectively, and choose to train the CIFAR10 using all these model architectures. For the \texttt{float32} model, we use typical training settings of \texttt{Relu} (BNN uses \texttt{HTanh}), ADAM optimizer, a batch size of 32 and a learning rate of $5\times 10^{-4}$. The \texttt{float32} model is trained with 100 epochs that is same when training BNN model.
Notably, this typical setting is not aiming to achieve state-of-the-art (SOTA) performance\cite{zheng2020learning}, which will require special training settings such as l2 regularization of the weights, warm-up and cosine annealing recession of the learning rate and powerful data enhancement methods, or even delicate loss functions\cite{zheng2020learning, cortes2012l2, loshchilov2016sgdr, cubuk2019autoaugment}.

\vspace{0.15cm}
\noindent{\bf A Comparison in Accuracy between Full-precision and 1-bit Model:} Results are detailed in Fig.~\ref{fig:Baseline}. It is expected that the BNN model accuracy is lower than the full precision model due to the trade-off between accuracy and precision~\cite{kundu2021dnr}. 
Notably, the focus of this work is not to optimize the BNN accuracy itself---for instance, one can exploit Neural
Architecture Search in the binary domain~\cite{bulat2020bats} for doing so; instead we aim to integrate multiple BNN tasks by sharing the same model parameter set. Our concern is whether adding more tasks will greatly affect the accuracy of the co-learned individual task. Once the baseline of the BNN accuracy is optimized, our RBNN will naturally inherit the accuracy optimization benefit.

\vspace{0.15cm}
\noindent{\bf Model Size Comparison between Full-precision and 1 bit Model:} The number of parameters of VGG, ResNet18 and ReActNet (ResNet14 as backbone) are 14.03MiB, 11.18MiB, and 2.78MiB, respectively\footnote{The model size is reported by \texttt{larq} library~\cite{larq}}. There is about $30\times$ model size reduction as a result of using the BNN model in comparison to its full precision model counterpart. For instance, the full precision model size of the VGG is up to 53.56MiB, while its BNN model size is only 1.74MiB. This demonstrates the advantage of employing BNN when the storage memory is limited.

\vspace{0.15cm}
\noindent{\bf Accuracy Baseline of Individual Task with 1 bit Model:} The baseline accuracy of each task individually trained by each of three models are summarized in Table~\ref{tab:baseline}. The testing accuracy is evaluated after 100 epochs of training---further increase in the number of epochs does not result in noticeable improvement in the accuracy in our experiments. The accuracy reported in Table~\ref{tab:baseline} serves as the baseline when comparing with the accuracy of each task in the RBNN when multiple tasks are concurrently trained. 

Here, we can see that the ReActNet has the best accuracy for the same task even though it is with the smallest model size. 
This is due to the fact that ReActNet is an optimized BNN framework that retains the full-precision representation for the fully-connect layers to maintain the model accuracy, in contrast the other two networks that use binary representation through the whole network. 

\begin{figure}[t]
	\centering
	\includegraphics[trim=0 0 0 0,clip,width=0.45\textwidth]{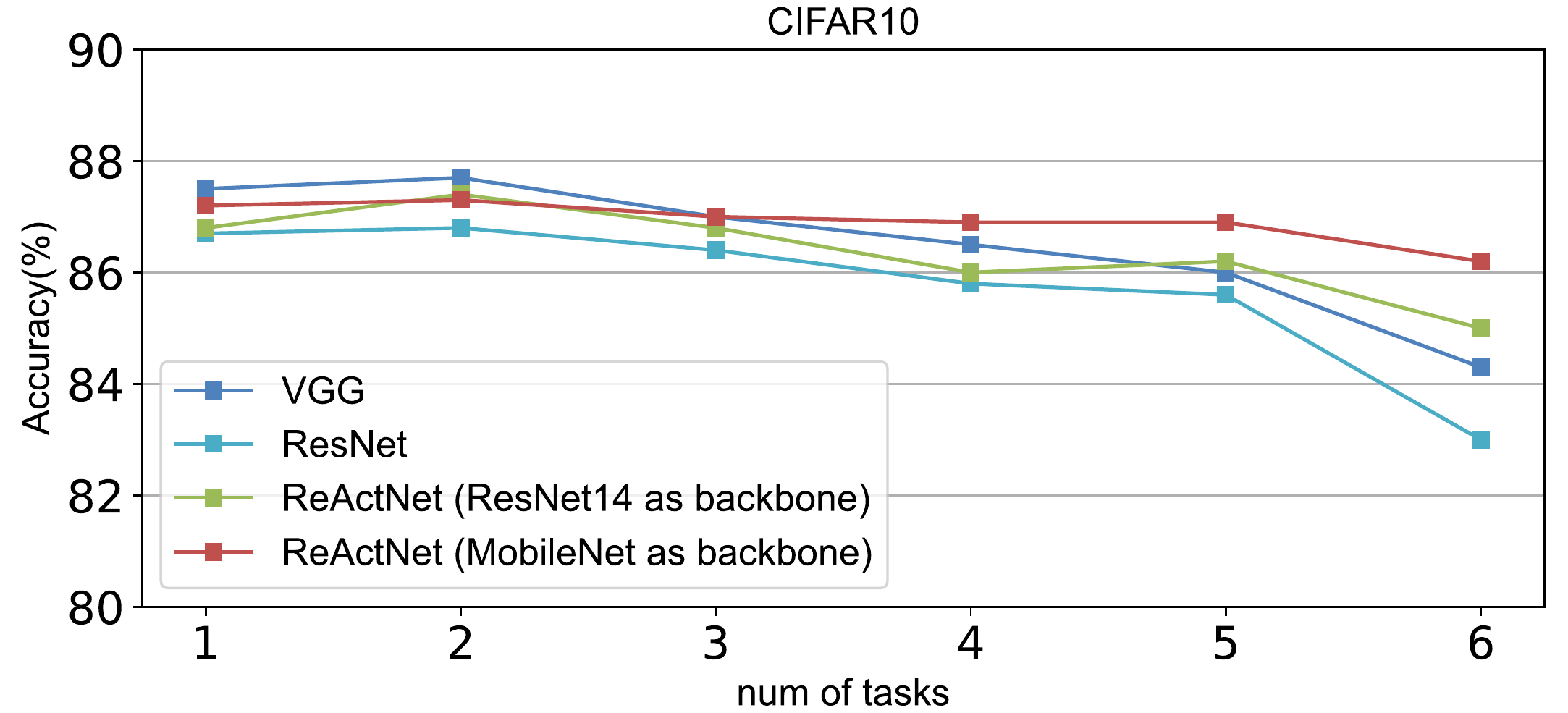}
	\caption{The accuracy of CIFAR10 task (it serves the first task integrated) when more other tasks are sequentially added into the RBNN.}
	\label{fig:CIFAR10Accuracy}
\end{figure}

\begin{table}
	\centering 
	\caption{Testing accuracy when each task/dataset is individually trained with a BNN model.}
			\resizebox{0.40\textwidth}{!}{
	\begin{tabular}{c| c | c} %
		\toprule 
		\toprule 
				
		Model Architect. & \begin{tabular}{@{}c@{}} Task/Dataset \end{tabular}   & \begin{tabular}{@{}c@{}} BNN Accuracy \\ (Baseline) \end{tabular} \\ 
		\toprule
	    \multirow{6}*{VGG}   & CIFAR10    & 87.49\%   \\
              & MNIST    & 99.21\%    \\
              & Fashion    & 94.57\%    \\
              & SVHN    & 95.18\%    \\
              & CelebA    & 97.25\%    \\
              & CIFAR100    & 57.30\%    \\
        \hline
	    \multirow{6}*{ResNet18}   & CIFAR10    & 86.69\%   \\
              & MNIST    & 99.12\%    \\
              & Fashion-MNIST    & 94.09\%    \\
              & SVHN    & 94.71\%    \\
              & CelebA    & 96.29\%    \\
              & CIFAR100    & 54.76\%    \\
        \hline
	    \multirow{7}*{ \begin{tabular}{@{}c@{}} ReActNet \\ (ResNet14 \\ as backbone) \end{tabular}}   & CIFAR10    & 86.81\%   \\
              & MNIST    & 99.31\%    \\
              & Fashion-MNIST    & 94.14\%    \\
              & SVHN    & 95.25\%    \\
              & CelebA    & 97.11\%    \\
              & CIFAR100    & 55.94\%    \\
              & SC    & 93.04\%    \\
              \hline
	    \multirow{7}*{ \begin{tabular}{@{}c@{}} ReActNet \\ (MobileNet \\ as backbone) \end{tabular}}   & CIFAR10    & 87.27\%   \\
              & MNIST    & 99.24\%    \\
              & Fashion-MNIST    & 94.19\%    \\
              & SVHN    & 95.08\%    \\
              & CelebA    & 97.26\%    \\
              & CIFAR100    & 57.63\%    \\
              & SC    & 93.21\%    \\
		\bottomrule
	\end{tabular}
			}
	\label{tab:baseline}
\end{table}

\begin{figure*}[t]
	\centering
	\includegraphics[trim=0 0 0 0,clip,width=0.95\textwidth]{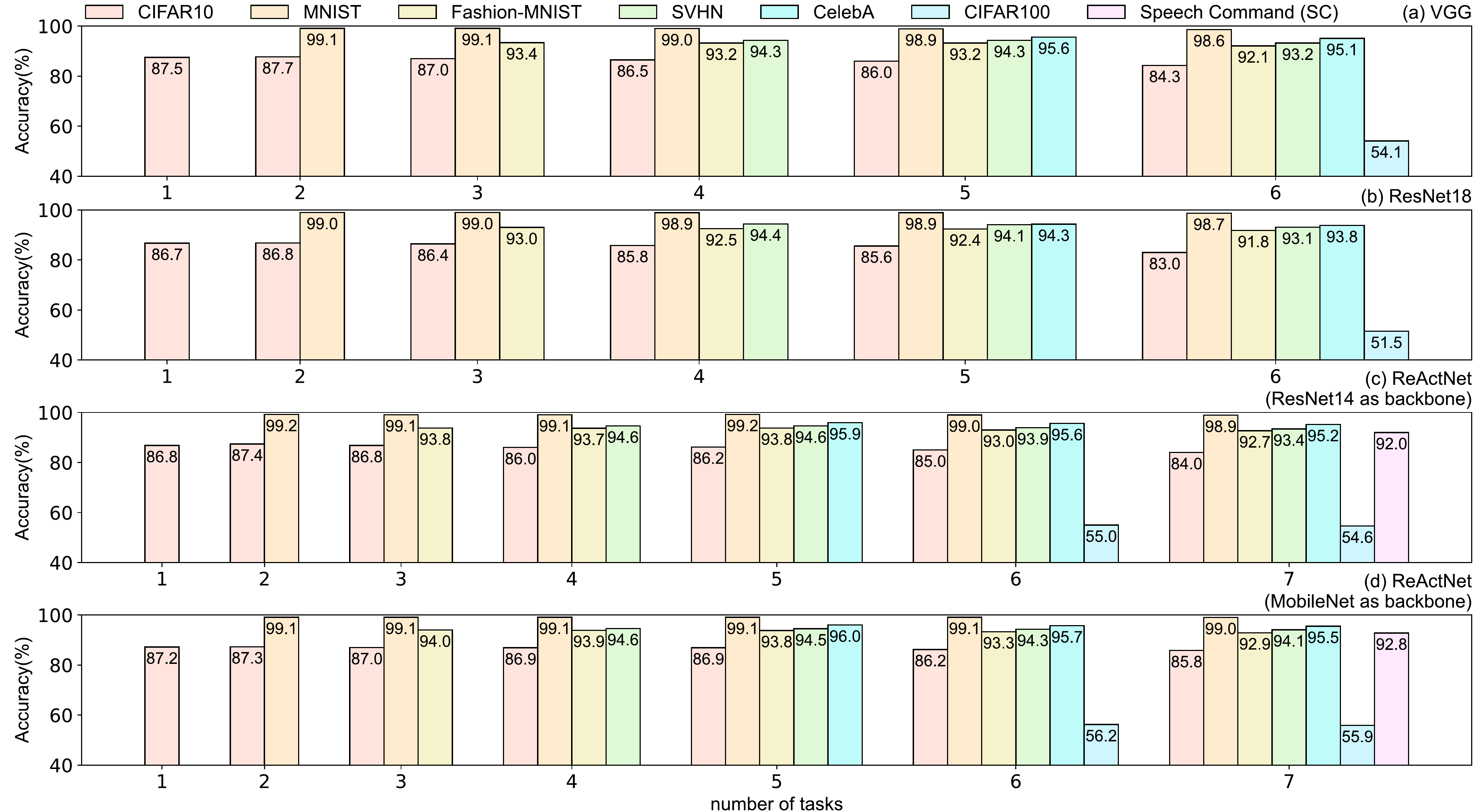}
	\caption{Testing accuracy of each individual task as the number of tasks increases. (a) VGG, (b) ResNet18, (c) ReActNet(ResNet14 as backbone), (d) ReActNet(MobileNet as backbone).}
	\label{fig:bar}
\end{figure*}

\subsubsection{RBNN Performance}
Here, we sequentially add individual tasks one-by-one into the RBNN to see how and to which extent the number of tasks will affect the accuracy of individual task. Comprehensive results are detailed in Fig.~\ref{fig:bar} for VGG, ResNet18, ReActNet (ResNet14 as backbone), and ReActNet (MobileNet as backbone), respectively. Corresponding to \textbf{Q1}, we have the following observations.

\begin{figure}[h]
	\centering
	\includegraphics[trim=0 0 0 0,clip,width=0.49\textwidth]{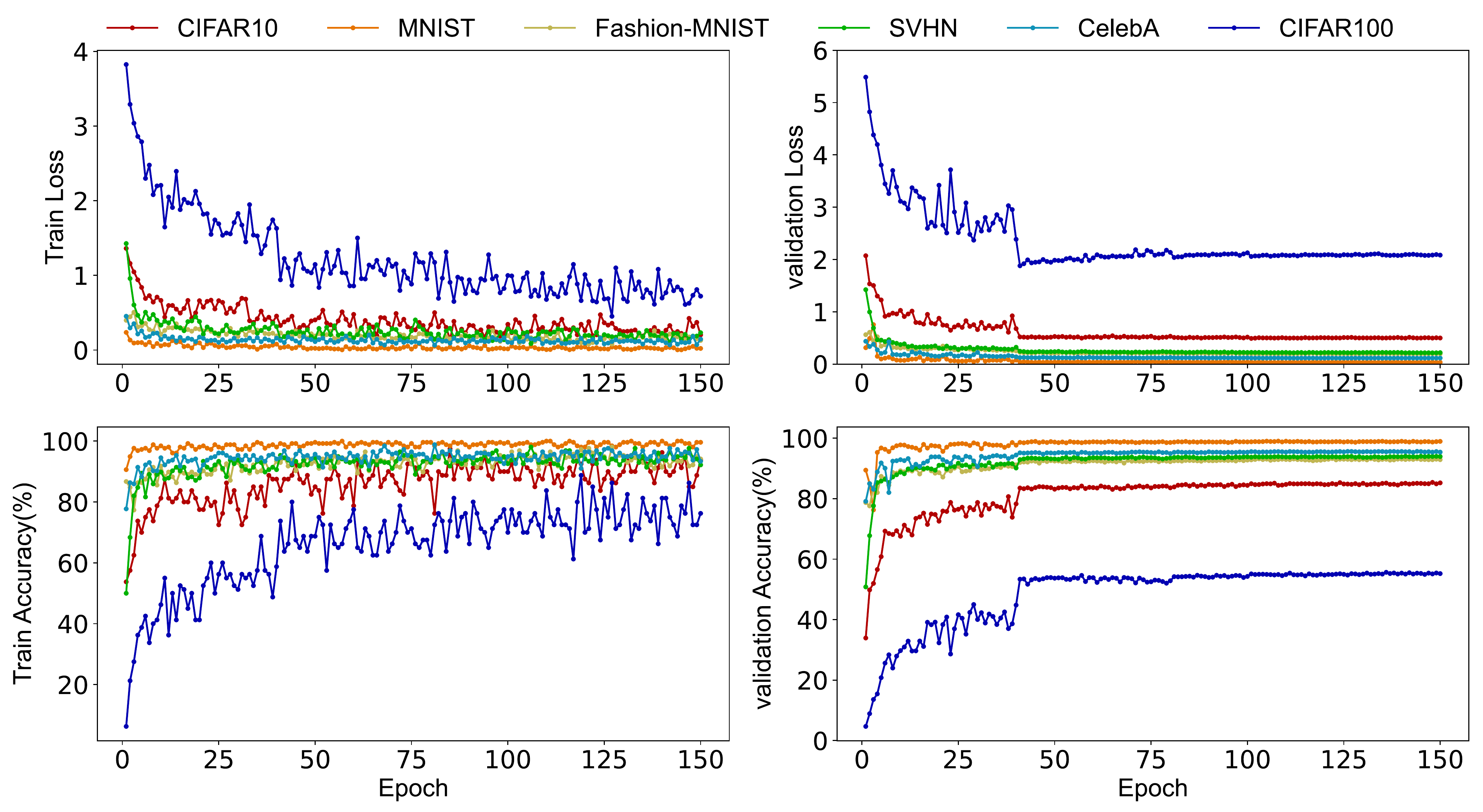}
	\caption{Loss and accuracy curves of each task during the RBNN training procedure. ReActNet with ResNet14 as backbone is used.}
	\label{fig:loss}
\end{figure}

Based on experiments with up to seven tasks, it is clear that multiple tasks can be embedded in the same BNN model, thus corroborating the RBNN. In addition, the accuracy of each task in the RBNN is similar to the trained alone task---this accuracy difference is less than 1\% in most cases. This is also exemplified in Fig.~\ref{fig:CIFAR10Accuracy} by the accuracy change of CIFAR10 task (it is the first task added) when more tasks are sequentially integrated. 
In addition to accuracy at the end of the RBNN training shown in Fig.~\ref{fig:bar}, the accuracy and loss changes per task during the RBNN training procedure are detailed in Fig.~\ref{fig:loss}. This further affirms \textbf{Q1}, where each task is learned concurrently and converged as the number of epochs increases---no over- or under-fitting is observed. Essentially, the loss and accuracy for each task converges in the RBNN training is in a similar trend when it is trained standalone. In addition, the loss and accuracy curves are experimental evidence that the catastrophic forgetting has been properly resolved in RBNN when multiple distinct tasks are co-learned.

As for RBNN with varying number of classes per task, it is still efficient. According to Table~\ref{tab:dataset}, CelebA task has two classes, the SC task has 35 classes, CIFAR100 task has 100 tasks, and all the rest four tasks have 10 classes. Therefore, the number of classes for each task varies greatly from 2 to 100. Overall, these tasks can co-exist with negligible accuracy drop. The worst accuracy drop occurs when the CIFAR100 is added, as it has the largest discrepancy with all other tasks in term of the number of classes. For instance, in the ResNet18, the CIFAR10 accuracy drops by about 2.8\% when the number of tasks increases from five to six (the sixth task is CIFAR100). At the meantime, Fashion-MNIST, SVHN see an accuracy drop of about 1\%. Evaluations on VGG and ReActNet model architectures demonstrate a similar tendency. Notably, the accuracy drop of other tasks when CIFAR100 is added is minimal for the ReActNet, the potential reason lies on the fact that ReActNet i) is delicately designed for BNN, and ii) has full-precision presentation of the full-connected layers. The later could be the main reason affording a better compatibility of tasks with varying number of classes. 

As for integrating tasks from different domains, we have added the SC task as an audio domain task that is distinct from image classification of all other six tasks. The results are shown in Fig.~\ref{fig:bar} using the ReActNet with ResNet14 and MobileNet as backbone, respectively. 
When SC is added, the accuracy drop of the other six image tasks is negligible, while the SC accuracy drop is about 1\% and thus also unnoticeable. Therefore, it is reasonable to conclude that the RBNN is able to learn tasks across different domains as long as the same model architecture can be used to these domains. This appears to be promising in practice as using same model architecture for cross-domain tasks is efficient. For instance, the \textit{transformer} model that achieves significant success in natural language processing, recently has demonstrated great potential in image-domain tasks~\cite{parmar2018image} and GAN-domain tasks~\cite{zhang2019self}.

Based on the experiments for three distinct model architectures, it is reasonable to conclude that the RBNN is generic to model architecture, regardless whether the model itself is specifically designed for BNN, which provides an affirmative answer to \textbf{Q1}. Noting the VGG and ResNet are not delicate for BNN, while only the ReActNet is.

\subsubsection{Increasing Quantization Width}

\begin{table}
	\centering 
	\caption{Testing accuracy when two tasks are integrated under varying quantization width.}
			\resizebox{0.45\textwidth}{!}{
	\begin{tabular}{c| c | c | c | c |} %
		\toprule 
		\toprule 
				
		Model & \begin{tabular}{@{}c@{}} Task \end{tabular} & \begin{tabular}{@{}c@{}} 1 bit \\ Quant. \\ Acc. \\ (B;R) \end{tabular} & \begin{tabular}{@{}c@{}} 4-bit \\ Quant. \\ Acc. \\ (B;R) \end{tabular} & \begin{tabular}{@{}c@{}} 8-bit \\ Quant. \\ Acc. \\ (B;R) \end{tabular} \\ 
		\toprule
	    \multirow{2}*{ResNet18}   & CIFAR10    & (86.69\%; 84.57\%) & (89.26\%; 88.42\%) & (88.92;88.62)   \\
              & CIFAR100    & (54.76\%; 54.55\%) & (58.07\%; 58.34\%) & (57.11\%; 57.76\%)   \\
		\bottomrule
	\end{tabular}
			}
	 \begin{tablenotes}
      \small
      \item (B;R): B means baseline accuracy when the task is \textit{individually} trained; R means when multiple tasks are \textit{co-learned}, standards for reconfiguration.
    \end{tablenotes}
	\label{tab:quantization} 
\end{table}

Though our main focus is BNN, here we extend the reconfiguration to other quantization width, in particularly, with 4-bit and 8-bit quantization\footnote{2-bit quantization exhibits unexpected lower accuracy, e.g., 50\% on CIFAR10 dataset trained by ResNet18. We note that our focus is BNN. This paragraph shows that our proposal is also applicable to other quantization width (including 2-bit quantization) for some device with relatively larger memory size.}. We use ResNet (in particular, ResNet18) with two tasks of CIFAR10 and CIFAR100 for the extensive experiments in this part. We note that CIFAR100 has the largest number of classes. All other settings including learning rate and number of epochs are similar to the one reported in Section~\ref{sec:modelArchitecture}.

As can be seen from Table~\ref{tab:quantization} the higher quantization width the better the accuracy of each task.
In our experiments, for both 4-bit and 8-bit quantizations, the CIFAR10 has been improved by more than 2\%. For CIFAR100, it increases by more than 3\%. In addition, we can see that the accuracy drop is minimized when a given task is co-learned in comparison with learned alone. 

This further validates the generic nature of the proposed reconfiguration, shows that  it is not solely limited to the concentrated BNN. Therefore, if the device has extra memory resources, then a slightly wider quantization approach can be used to improve the accuracy.
Therefore, if the device has extra memory resources, then a slightly wider quantization approach can be used to improve the accuracy. 

\section{IP protection}\label{sec:IPProtection}

It is recognized that the reconfiguration naturally enables model IP protection, e.g., allowing a user for trial and to only pay per the task needed (\textbf{Q2}). For instance, the default task is provided to the user for trial before payment. The trial task has partial functionality, e.g., classifying a limited number of classes. Once the user is satisfied with the trial and pays for the license, the model provider sends a key to the user, ${\rm User\_key}_M$, to reconfigure and thus unlock the task with a full functionality. Notably, we have shown that tasks with varying number of classes can be effectively co-existed in RBNN. This scheme facilities the user experience while protecting model provider IP. As for the user, she only needs to pay the task $T_m$ that she is interested to save cost. Next we propose a scheme to allow the user pay not only per task but also per device.

\begin{figure}[t]
	\centering
	\includegraphics[trim=0 0 0 0,clip,width=0.35\textwidth]{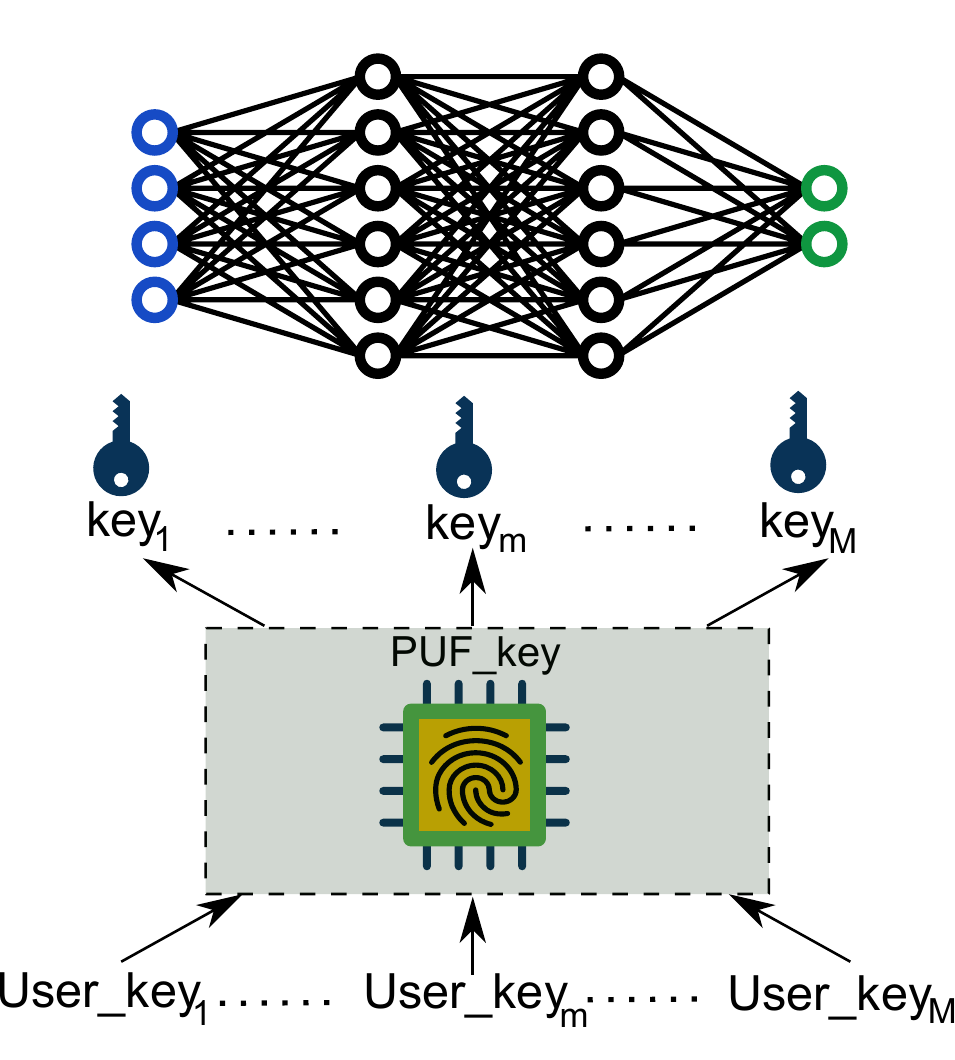}
	\caption{The ${\rm key}_M$ that reconfigures the model into the $M_{\rm th}$ task is a function of both the ${\rm User\_key}_M$ and the PUF\_key. The PUF\_key is extracted from the PUF alike `fingerprint', and thus uniquely bounds up with each hardware device.}
	\label{fig:PUFkey}
\end{figure}

Here, we propose to bind the key reconfiguring a different task with a unique device key that is extracted from the physical unclonable function (PUF)~\cite{gao2020physical,gu2019flip}. 
In general, the PUF is akin the \textit{fingerprint} of the device, such a fingerprint can be converted into a usable cryptographic key, PUF\_key. The PUF is a result of inevitable and uncontrollable manufacturing variations, thus preventing forging two identical PUF instances even under the same manufacturing process. Thus, the PUF\_key is uniquely bound up with \textit{each device}, which is hard to be physically cloned and replaced. Here, we can consider employing the intrinsic SRAM PUF to extract the PUF\_key, since the SRAM is pervasively embedded within commodity electronics, requiring no additional hardware modification. The details of extracting PUF\_key from (even low-end) commodity electronics in a lightweight manner can be referred to~\cite{gao2018lightweight}.

One prominent advantage of this PUF\_key enabled reconfiguration is that the payment now can be made per device, and even per task. For the former, the model provider can use the same ${\rm key}_m$ to reconfigure the same task for avoiding training the RBNN per device. From the user's perspective, the ${\rm User\_key}_m$ is still different per device as the ${\rm key}_m$ is also a function of the PUF\_key that is unknown, as illustrated in Fig.~\ref{fig:PUFkey}. Thus, the provider can issue different ${\rm User\_key}_m$ to different device given the same task. Here, we can simply set ${\rm key}_m$$=$\textsf{XOR}(${\rm User\_key}_m$, PUF\_key)---admittedly, other means of deriving ${\rm key}_m$ upon ${\rm User\_key}_m$ and PUF\_key can be adopted.

As for the latter, the user can choose to only pay for a specific task $T_m$. If later another task is needed, the user can request from the model provider to issue another ${\rm User\_key_i}$ for the task of interest. In either case, the issued ${\rm User\_key_m}$ can not be used on other devices even if every device has been provided all tasks by the model provider. However, without knowing the ${\rm User\_key_m}$ that is specific to the device in-hand, the user cannot activate the task on the device, which protects the IP of the provider not only per device but also per task. At the same time, this allows the user to only pay for the task of interest on that device. 

\section{Comparisons and Discussion}\label{sec:CompareAndDiscuss}
We compare the RBNN with other approaches and provide further discussions on such as its insensitivity to task adding order and adaption of activation functions.
\subsection{Task Integration Comparison}

\begin{table*}[h]
\centering 
\caption{An accuracy comparison of multi-task learning between merged datasets and RBNN.}
\resizebox{0.9 \textwidth}{!}
{
\begin{tabular}{c | c | c | c | c | c | c | c} %
\toprule 
Order & Methods & $\rm CIFAR10^{1}$ & $\rm MNIST^{2}$  & $\rm Fashion-MNIST^{3}$ & $\rm SVHN^{4}$ & $\rm CelebA^{5}$ & $\rm CIFAR100^{6}$ \\ 

\midrule
- & BNN(individually trained) &
86.8\% & 99.3\%  & 94.1\% & 95.3\% & 97.1\% & 55.9\% \\

\midrule
\multirow{2}*{(1, 2, 3, 4)}
& Merged datasets &  85.4\% & 99.0\%  & 93.3\% & 93.7\% & - & - \\
& RBNN  &  86.0\% & 99.1\%  & 93.7\% & 94.6\% & - & - \\

\midrule
\multirow{2}*{(1, 2, 3, 4, 5)}
& Merged datasets &  85.2\% & 98.9\%  & 93.4\% & 93.5\% & 95.6\% & - \\
& RBNN  &  86.2\% & 99.2\%  & 93.8\% & 94.6\% & 95.9\% & - \\

\midrule
\multirow{2}*{(1, 2, 3, 4, 5, 6)}
& Merged datasets & 79.7\% & 98.7\%  & 92.2\% & 92.0\% & 95.5\% & 52.9\% \\
& RBNN  &  85.0\% & 99.0\%  & 93.0\% & 93.9\% & 95.6\% & 55.0\% \\

\bottomrule
\end{tabular}
}
	 \begin{tablenotes}
      \small
      \item The superscript number in e.g., $\rm CIFAR10^{1}$ means the number denotation of the task for an abbreviated reference.
    \end{tablenotes}
\label{tab:integrationbaseline}
\end{table*}

As the RBNN is the first work demonstrating the feasibility of integrating multiple tasks into a single BNN model to reduce the DNN required memory footprint substantially, there is no other explicitly related work similar to our study.

Nonetheless, we have used the other means of integrating multiple tasks into the same model to serve as a multiple task integration baseline model, which generally merges these tasks into one and trains the model---thus, different tasks sharing similar latent representations. In this new baseline experiments, we merge 4 to 6 datasets and train the merged large dataset on ReActNet (ResNet14 as backbone) as the ReActNet is a delicate design for BNN. The accuracy of each merged task is (notably) smaller than that of RBNN (could be lower by 5.3\%) as shown in Table~\ref{tab:integrationbaseline}. The difference in the inference accuracy between the merged tasks and the RBNN increases slightly when more datasets are merged, especially for datasets with feature similarity (we emphasize that CIFAR10 and CIFAR100 have no overlapped categories). More importantly, merging disparate data sets into a larger one and training with the same model architecture will \textit{no longer provide desirable IP protection as the RBNN does}.

\subsection{Continual Learning}
Continual learning (CL), or long-life learning, may possibly an alternative that promises to reduce the memory requirement when it is incorporated into binary neural network, which we have not explored in this study. This is an open question for future work. Interestingly, we find that the techniques of isolating parameters per task exhibiting rigorous performance~\cite{delange2021continual} in CL somehow shares similar functionalities of the parameter permutation used in RBNN, where the tasks do not share same latent representation to deal with forgetting. In addition, the insensitivity of task order in CL~\cite{delange2021continual} aligns our validations of RBNN that is insensitive to task order.

Nonetheless, the aims and learning processes of CL and RBNN are distinct. Note that CL has a (dynamic) sequential learning process to accumulate new knowledge when new data of new task comes. However, the data of previous task(s)---the future task may also be unknown---is unavailable or only limited data samples are accessible. In other words, a major criterion of CL is the sequential nature of the learning process, where only a small portion of input data from one or few tasks are available at once~\cite{delange2021continual}. In contrast, the RBNN has no such premise, as it trains multiple tasks with access to all training data at once during the co-learning. 
Once these multiple tasks are learned, they will not be updated by design since they will be sold as IP and deployed on IoT devices. 
In the CL, the semantics of the tasks play an important role in the amount of forgetting~\cite{ramasesh2020anatomy}. Nonetheless, catastrophic forgetting in CL is challenging but can be circumvented with special measures~\cite{delange2021continual}. For RBNN, it does not need the semantics of the tasks (similarities/dissimilarities among tasks), as  it integrates multiple models without sharing any latent representations among tasks. This enables the reconfigurability to be bound with a distinct key, thus protecting the model IP.

More specifically, the aims of RBNN are to reduce the memory and computation overhead when the deep learning model is deployed on mobile devices by switching inference tasks on demand. The RBNN has no any prior (semantic) requirement on the integrated tasks. The CL is a general approach to accumulate new knowledge sequentially, which can learn multiple tasks but appear to have stringent requirements on the semantics among tasks. The CL may be a promising alternative to reduce the memory and computation overhead by incorporating it with BNN. In addition, the RBNN is devised to provide IP protections by simply binding the task switch with a key, whereas the CL is unable to provide such a security gain, at least trivially.

\subsection{Model Obfuscation Comparison}
Chakraborty~\textit{et al.}~\cite{chakraborty2020hardware} recently proposed the first work to leverage hardware as a root-of-trust to achieve IP security of DNN models, which is to obfuscate the model parameters in a way that the model is only correctly functioning on an authorized device with the unlocking key. From the security perspective of locking the DNN model (notably, security is one of two main benefits of our proposed RBNN), it is related to our work. From the \textit{security perspective}, we differentiate our work from~\cite{chakraborty2020hardware} in three main aspects:
\begin{itemize}
    \item Generally, in~\cite{chakraborty2020hardware}, the output of an internal neuron is now \textit{a function of both its original input and the hardware provided key}---so bringing obfuscation for locking. The RBNN model locking technique is different in a way that we leverage \textit{parameter permutation} in the RBNN training as a means of obfuscation---this is a byproduct of the RBNN to reduce memory footprint.
    \item In contrast to the assumption in~\cite{chakraborty2020hardware} that the key is securely stored in a secure storage in, e.g., Trusted Platform Module (TPM), we do not require the secure key storage, as secure storage is unlikely available for a range of medium/low-end mobile devices. We propose to utilize the SRAM PUF that is pervasively used in electronic devices to provide the unique key bound up with the hardware. This indicates that our solution can be applied to a wider range of electronic computing devices.
    \item During the training of the locked model in~\cite{chakraborty2020hardware}, the key value must be known. In our case, we do not need this prerequisite. More specifically, an RBNN task-specific ${\rm key}_m$ is a function of ${\rm User}\_{\rm key}_m$ and a device-unique ${\rm PUF\_key}$. The ${\rm key}_m$ can be arbitrarily selected but kept same (note it is hidden from the user) among devices. Then the user key ${\rm User}\_{\rm key}_m$ is authorized according to the device ${\rm PUF\_key}$, so that the user key is ensured not user-specific but also device specific.
\end{itemize}

\subsection{Task Adding Order}\label{sec:taskadding}

\begin{table*}
\centering 
\caption{The effect of task adding order on model performance.}
\resizebox{0.75 \textwidth}{!}
{
\begin{tabular}{c | c | c | c | c | c | c} %
\toprule 
Order & $\rm CIFAR10^{1}$ & $\rm MNIST^{2}$  & \begin{tabular}{@{}c@{}} Fashion-MNIST$^{3}$ \end{tabular}  & $\rm SVHN^{4}$ & $\rm CelebA^{5}$ & $\rm CIFAR100^{6}$ \\ 

\midrule
(1, 2, 3, 4, 5, 6) &
85.0\% & 99.0\%  & 93.0\% & 93.9\% & 95.6\% & 55.0\% \\

\midrule
(6, 5, 4, 3, 2, 1) &
84.9\% & 99.0\%  & 93.1\% & 93.7\% & 95.7\% & 55.0\% \\

\bottomrule
\end{tabular}
}
    \begin{tablenotes}
      \small
      \item The superscript number in, e.g., $\rm CIFAR10^{1}$ means the number denotation of the task for an abbreviated reference.
    \end{tablenotes}
\label{tab:order}
\end{table*}

Based on Eq.~\ref{eq:loss} where sub-loss of each task has an independent addition relationship, RBNN is insensitive to the adding order in principle. In addition, further experimental results are shown in Table~\ref{tab:order} with a reverse order, which do confirm the RBNN's invariant to the task adding order. As we can clearly observe that the accuracy of each task before and after re-ordering are almost the same with negligible variations caused by rerunning the model inference. More generally, the dataset of each task is used to update the reconfigured weights corresponding to the task independently. As long as different tasks are alternately trained in the same order for each batch unit and the order can be randomly initialized, the same performance of the RBNN can be ensured.

\subsection{Hard-to-Learn Task}
In principle, the RBNN scales to more difficult tasks. If the hard-to-learn task's input size is larger and number of classes are higher than the tasks tested in this work, a more complex underlying RBNN model is required. In other words, the hard-to-learn task itself could require a deeper model with larger input size to gain a better accuracy. As a consequence, the complexity of the underlying RBNN model should also be increased as a prerequisite.

\subsection{Activation Function}
RBNN by design is independent of the activation function such as \textsf{Htanh}, \textsf{softplus} and \textsf{Relu}. However, \textsf{Htanh} is more suitable for BNN since it naturally limits the activation output to between -1 and 1. This is the reason we chose to use \textsf{Htanh} in our RBNN realization as in Eq.~\ref{eq:htanh}. We now evaluate the RBNN performance when activation functions other than \textsf{Htanh} are adopted. Experiments are implemented on the ReActNet model (ResNet14 as backbone), where six subtasks were co-trained by the RBNN. The results are summarized in Table~\ref{tab:activation}. The inference accuracy for each subtask except CIFAR100 was roughly equivalent when using different activation functions---\textsf{Htanh} is slightly better. For CIFAR100, \textsf{Htanh} (55\%) is relatively higher than \textsf{softplus} (53.2\%) and \textsf{Relu} (using \textsf{ReLU6}) (52.9\%). A possible reason for this is that neither \textsf{Relu} nor \textsf{softplus} support negative values (e.g., \textsf{Relu} is truncated to zero---\textsf{softplus} is a smoothed version of \textsf{Relu}).

\begin{table*}
\centering 
\caption{The effect of activation function on model performance.}
\resizebox{0.7 \textwidth}{!}
{
\begin{tabular}{c | c | c | c | c | c | c} %
\toprule 
Activation function & CIFAR10 & MNIST  & \begin{tabular}{@{}c@{}} Fashion-MNIST \end{tabular}  & SVHN & CelebA & CIFAR100 \\ 

\midrule
\textsf{softplus} &
84.2\% & 99.0\%  & 92.7\% & 93.7\% & 95.5\% & 53.2\% \\

\midrule
\textsf{ReLU6} &
84.1\% & 98.8\%  & 92.7\% & 93.6\% & 95.4\% & 52.9\% \\

\midrule
\textsf{HTanh} &
85.0\% & 99.0\%  & 93.0\% & 93.9\% & 95.6\% & 55.0\% \\

\bottomrule
\end{tabular}
}
\label{tab:activation}
\end{table*}

\subsection{Limitations}

The RBNN reconfiguration on-device is mainly determined by feeding a task-specific ${\rm key}_m$, with ${\rm key}_m$ a function of ${\rm User}\_{\rm key}_m$ and a device-unique ${\rm PUF\_key}$. As the ${\rm User}\_{\rm key}_m$ is provided to the user once the user buys the IP, the ${\rm PUF\_key}$ requires online derivation that can dominate the reconfiguration overhead. Based on a recent work~\cite{gao2021noisfre}, it has been shown that the pervasively electronic commodities integrated intrinsic SRAM PUF based ${\rm PUF\_key}$ can be securely provided with substantially reduced overhead. Specifically, it requires 44,082 clock cycles for a 128-bit ${\rm PUF\_key}$ derivation when the realization is performed on a low-end nRF52832 chip---a popular RF-enabled MCU. Such small overhead appears not a concern for mobile devices or medium-end IoT devices that the RBNN target.

Nonetheless, as this work mainly focuses on proposing and validating a new paradigm for reducing the DNN model's memory volume requirement by the RBNN and a framework to protect the DNN model IP naturally enabled by the RBNN, the overhead of end-to-end RBNN realization on a dedicated hardware or a microcontroller is not specifically evaluated. This is a limitation of present study, which is an open question for future work.

\section{Conclusion} \label{sec:Conclusion}

We have proposed RBNN to further improve the memory utilization for resource-constrained IoT devices by a factor of $M$, with $M$ as the number of tasks concurrently embedded. Note that RBNN is built upon the ultimate binary quantization of DNN---the reconfigurable paradigm indeed does apply to non-binary neural networks. Comprehensive experiments validate the practicability and effectiveness of the RBNN by: i) up to seven tasks trained over each of three common model architectures; ii) tasks with varying number of classes; iii) tasks spanning distinct domains. 
More specifically, the task accuracy in the RBNN has no or only negligible degradation in comparison with standalone trained BNN tasks. In addition, the RBNN naturally enables novel IP protection for the model provider. The IP of each task integrated within RBNN can even be issued per user as well as per device, which greatly reduces the IP purchase for consumers while reserving the commercial benefits for the model provider.

\bibliographystyle{IEEEtran}
\bibliography{References}

\begin{thebibliography}{10}
\providecommand{\url}[1]{#1}
\csname url@samestyle\endcsname
\providecommand{\newblock}{\relax}
\providecommand{\bibinfo}[2]{#2}
\providecommand{\BIBentrySTDinterwordspacing}{\spaceskip=0pt\relax}
\providecommand{\BIBentryALTinterwordstretchfactor}{4}
\providecommand{\BIBentryALTinterwordspacing}{\spaceskip=\fontdimen2\font plus
\BIBentryALTinterwordstretchfactor\fontdimen3\font minus
  \fontdimen4\font\relax}
\providecommand{\BIBforeignlanguage}[2]{{%
\expandafter\ifx\csname l@#1\endcsname\relax
\typeout{** WARNING: IEEEtran.bst: No hyphenation pattern has been}%
\typeout{** loaded for the language `#1'. Using the pattern for}%
\typeout{** the default language instead.}%
\else
\language=\csname l@#1\endcsname
\fi
#2}}
\providecommand{\BIBdecl}{\relax}
\BIBdecl

\bibitem{boulemtafes2020review}
A.~Boulemtafes, A.~Derhab, and Y.~Challal, ``A review of privacy-preserving
  techniques for deep learning,'' \emph{Neurocomputing}, vol. 384, pp. 21--45,
  2020.

\bibitem{kundu2021analyzing}
S.~Kundu, Q.~Sun, Y.~Fu, M.~Pedram, and P.~Beerel, ``Analyzing the
  confidentiality of undistillable teachers in knowledge distillation,''
  \emph{Advances in Neural Information Processing Systems}, vol.~34, 2021.

\bibitem{krizhevsky2017imagenet}
A.~Krizhevsky, I.~Sutskever, and G.~E. Hinton, ``Imagenet classification with
  deep convolutional neural networks,'' \emph{Communications of the ACM},
  vol.~60, no.~6, pp. 84--90, 2017.

\bibitem{iandola2016squeezenet}
F.~N. Iandola, S.~Han, M.~W. Moskewicz, K.~Ashraf, W.~J. Dally, and K.~Keutzer,
  ``Squeezenet: Alexnet-level accuracy with 50x fewer parameters and {$<$0.5
  MB} model size,'' \emph{arXiv preprint arXiv:1602.07360}, 2016.

\bibitem{he2016deep}
K.~He, X.~Zhang, S.~Ren, and J.~Sun, ``Deep residual learning for image
  recognition,'' in \emph{Proceedings of the IEEE Conference on Computer Vision
  and Pattern Recognition}, 2016, pp. 770--778.

\bibitem{han2020ghostnet}
K.~Han, Y.~Wang, Q.~Tian, J.~Guo, C.~Xu, and C.~Xu, ``Ghostnet: More features
  from cheap operations,'' in \emph{Proceedings of the IEEE Conference on
  Computer Vision and Pattern Recognition}, 2020, pp. 1580--1589.

\bibitem{zhang2018shufflenet}
X.~Zhang, X.~Zhou, M.~Lin, and J.~Sun, ``Shufflenet: An extremely efficient
  convolutional neural network for mobile devices,'' in \emph{Proceedings of
  the IEEE Conference on Computer Vision and Pattern Recognition}, 2018, pp.
  6848--6856.

\bibitem{david2020tensorflow}
R.~David, J.~Duke, A.~Jain, V.~J. Reddi, N.~Jeffries, J.~Li, N.~Kreeger,
  I.~Nappier, M.~Natraj, S.~Regev \emph{et~al.}, ``{TensorFlow Lite Micro}:
  Embedded machine learning on tinyml systems,'' \emph{arXiv preprint
  arXiv:2010.08678}, 2020.

\bibitem{TensorflowLite}
\BIBentryALTinterwordspacing
(Oct 2021) Tensorflow lite. [Online]. Available:
  \url{https://www.tensorflow.org/lite}
\BIBentrySTDinterwordspacing

\bibitem{hubara2016binarized}
I.~Hubara, M.~Courbariaux, D.~Soudry, R.~El-Yaniv, and Y.~Bengio, ``Binarized
  neural networks,'' \emph{Advances in Neural Information Processing Systems},
  vol.~29, pp. 4107--4115, 2016.

\bibitem{rastegari2016xnor}
M.~Rastegari, V.~Ordonez, J.~Redmon, and A.~Farhadi, ``Xnor-net: Imagenet
  classification using binary convolutional neural networks,'' in
  \emph{European Conference on Computer Vision}.\hskip 1em plus 0.5em minus
  0.4em\relax Springer, 2016, pp. 525--542.

\bibitem{bulat2019xnor}
\BIBentryALTinterwordspacing
A.~Bulat and G.~Tzimiropoulos, ``{XNOR}-net++: Improved binary neural
  networks,'' in \emph{British Machine Vision Conference (BMVC)}, 2019.
  [Online]. Available: \url{https://arxiv.org/abs/1909.13863}
\BIBentrySTDinterwordspacing

\bibitem{martinez2020training}
\BIBentryALTinterwordspacing
B.~Martinez, J.~Yang, A.~Bulat, and G.~Tzimiropoulos, ``Training binary neural
  networks with real-to-binary convolutions,'' in \emph{International
  Conference on Learning Representations (ICLR)}, 2020. [Online]. Available:
  \url{https://arxiv.org/abs/2003.11535}
\BIBentrySTDinterwordspacing

\bibitem{andri2017yodann}
R.~Andri, L.~Cavigelli, D.~Rossi, and L.~Benini, ``Yodann: An architecture for
  ultralow power binary-weight {CNN} acceleration,'' \emph{IEEE Transactions on
  Computer-Aided Design of Integrated Circuits and Systems}, vol.~37, no.~1,
  pp. 48--60, 2017.

\bibitem{conti2018xnor}
F.~Conti, P.~D. Schiavone, and L.~Benini, ``{XNOR} neural engine: A hardware
  accelerator {IP} for 21.6-fj/op binary neural network inference,'' \emph{IEEE
  Transactions on Computer-Aided Design of Integrated Circuits and Systems},
  vol.~37, no.~11, pp. 2940--2951, 2018.

\bibitem{qin2020binary}
H.~Qin, R.~Gong, X.~Liu, X.~Bai, J.~Song, and N.~Sebe, ``Binary neural
  networks: A survey,'' \emph{Pattern Recognition}, p. 107281, 2020.

\bibitem{zhang2020fracbnn}
Y.~Zhang, J.~Pan, X.~Liu, H.~Chen, D.~Chen, and Z.~Zhang, ``{FracBNN}: Accurate
  and {FPGA}-efficient binary neural networks with fractional activations,'' in
  \emph{The 2021 ACM/SIGDA International Symposium on Field-Programmable Gate
  Arrays}, 2021, pp. 171--182.

\bibitem{adi2018turning}
Y.~Adi, C.~Baum, M.~Cisse, B.~Pinkas, and J.~Keshet, ``Turning your weakness
  into a strength: Watermarking deep neural networks by backdooring,'' in
  \emph{27th USENIX Security Symposium (USENIX Security 18)}, 2018, pp.
  1615--1631.

\bibitem{darvish2019deepsigns}
B.~Darvish~Rouhani, H.~Chen, and F.~Koushanfar, ``Deepsigns: An end-to-end
  watermarking framework for ownership protection of deep neural networks,'' in
  \emph{Proceedings of the Twenty-Fourth International Conference on
  Architectural Support for Programming Languages and Operating Systems}, 2019,
  pp. 485--497.

\bibitem{lin2020chaotic}
N.~Lin, X.~Chen, H.~Lu, and X.~Li, ``Chaotic weights: A novel approach to
  protect intellectual property of deep neural networks,'' \emph{IEEE
  Transactions on Computer-Aided Design of Integrated Circuits and Systems,
  DOI: 10.1109/TCAD.2020.3018403}, 2020.

\bibitem{chakraborty2020hardware}
A.~Chakraborty, A.~Mondai, and A.~Srivastava, ``Hardware-assisted intellectual
  property protection of deep learning models,'' in \emph{2020 57th ACM/IEEE
  Design Automation Conference (DAC)}.\hskip 1em plus 0.5em minus 0.4em\relax
  IEEE, 2020, pp. 1--6.

\bibitem{ruder2017overview}
S.~Ruder, ``An overview of multi-task learning in deep neural networks,''
  \emph{arXiv preprint arXiv:1706.05098}, 2017.

\bibitem{gao2019strip}
Y.~Gao, C.~Xu, D.~Wang, S.~Chen, D.~C. Ranasinghe, and S.~Nepal, ``Strip: A
  defence against trojan attacks on deep neural networks,'' in
  \emph{Proceedings of the 35th Annual Computer Security Applications
  Conference}, 2019, pp. 113--125.

\bibitem{han2015deep}
\BIBentryALTinterwordspacing
S.~Han, H.~Mao, and W.~J. Dally, ``Deep compression: Compressing deep neural
  networks with pruning, trained quantization and {Huffman} coding,'' in
  \emph{NIPS Deep Learning Symposium}, 2015. [Online]. Available:
  \url{https://arxiv.org/abs/1510.00149}
\BIBentrySTDinterwordspacing

\bibitem{kundu2021dnr}
S.~Kundu, M.~Nazemi, P.~A. Beerel, and M.~Pedram, ``{DNR}: A tunable robust
  pruning framework through dynamic network rewiring of {DNN}s,'' in
  \emph{Proceedings of the 26th Asia and South Pacific Design Automation
  Conference}, 2021, pp. 344--350.

\bibitem{hinton2015distilling}
G.~Hinton, O.~Vinyals, and J.~Dean, ``{Distilling the Knowledge in a Neural
  Network},'' \emph{arXiv preprint arXiv:1503.02531}, 2015.

\bibitem{polino2018model}
\BIBentryALTinterwordspacing
A.~Polino, R.~Pascanu, and D.~Alistarh, ``Model compression via distillation
  and quantization,'' in \emph{International Conference on Learning
  Representations (ICLR)}, 2018. [Online]. Available:
  \url{https://arxiv.org/abs/1802.05668}
\BIBentrySTDinterwordspacing

\bibitem{kundu2022bmpq}
S.~Kundu, S.~Wang, Q.~Sun, P.~A. Beerel, and M.~Pedram, ``{BMPQ}: bit-gradient
  sensitivity-driven mixed-precision quantization of {DNN}s from scratch,'' in
  \emph{2022 Design, Automation \& Test in Europe Conference \& Exhibition
  (DATE)}.\hskip 1em plus 0.5em minus 0.4em\relax IEEE, 2022, pp. 588--591.

\bibitem{xu2018deepobfuscation}
H.~Xu, Y.~Su, Z.~Zhao, Y.~Zhou, M.~R. Lyu, and I.~King, ``Deepobfuscation:
  Securing the structure of convolutional neural networks via knowledge
  distillation,'' \emph{arXiv preprint arXiv:1806.10313}, 2018.

\bibitem{gao2020backdoor}
Y.~Gao, B.~G. Doan, Z.~Zhang, S.~Ma, A.~Fu, S.~Nepal, and H.~Kim, ``Backdoor
  attacks and countermeasures on deep learning: a comprehensive review,''
  \emph{arXiv preprint arXiv:2007.10760}, 2020.

\bibitem{bagdasaryan2020blind}
\BIBentryALTinterwordspacing
E.~Bagdasaryan and V.~Shmatikov, ``Blind backdoors in deep learning models,''
  in \emph{USENIX Security Symposium}, 2020. [Online]. Available:
  \url{https://arxiv.org/abs/2005.03823}
\BIBentrySTDinterwordspacing

\bibitem{guo2020trojannet}
\BIBentryALTinterwordspacing
C.~Guo, R.~Wu, and K.~Q. Weinberger, ``Trojannet: Exposing the danger of trojan
  horse attack on neural networks,'' in \emph{The International Conference on
  Learning Representations (ICLR)}, 2020. [Online]. Available:
  \url{https://openreview.net/forum?id=BJeGA6VtPS}
\BIBentrySTDinterwordspacing

\bibitem{bagdasaryan2020backdoor}
E.~Bagdasaryan, A.~Veit, Y.~Hua, D.~Estrin, and V.~Shmatikov, ``How to backdoor
  federated learning,'' in \emph{International Conference on Artificial
  Intelligence and Statistics}.\hskip 1em plus 0.5em minus 0.4em\relax PMLR,
  2020, pp. 2938--2948.

\bibitem{shan2020gotta}
S.~Shan, E.~Wenger, B.~Wang, B.~Li, H.~Zheng, and B.~Y. Zhao, ``Gotta catch'em
  all: Using honeypots to catch adversarial attacks on neural networks,'' in
  \emph{Proceedings of the ACM SIGSAC Conference on Computer and Communications
  Security}, 2020, pp. 67--83.

\bibitem{sommer2020towards}
D.~M. Sommer, L.~Song, S.~Wagh, and P.~Mittal, ``Towards probabilistic
  verification of machine unlearning,'' \emph{arXiv preprint arXiv:2003.04247},
  2020.

\bibitem{kirkpatrick2017overcoming}
J.~Kirkpatrick, R.~Pascanu, N.~Rabinowitz, J.~Veness, G.~Desjardins, A.~A.
  Rusu, K.~Milan, J.~Quan, T.~Ramalho, A.~Grabska-Barwinska \emph{et~al.},
  ``Overcoming catastrophic forgetting in neural networks,'' \emph{Proceedings
  of the National Academy of Sciences}, vol. 114, no.~13, pp. 3521--3526, 2017.

\bibitem{hinton2012neural}
G.~Hinton, N.~Srivastava, and K.~Swersky, ``Neural networks for machine
  learning,'' \emph{Coursera, video lectures}, vol. 264, no.~1, 2012.

\bibitem{lecun1998gradient}
Y.~LeCun, L.~Bottou, Y.~Bengio, and P.~Haffner, ``Gradient-based learning
  applied to document recognition,'' \emph{Proceedings of the IEEE}, vol.~86,
  no.~11, pp. 2278--2324, 1998.

\bibitem{xiao2017fashion}
H.~Xiao, K.~Rasul, and R.~Vollgraf, ``Fashion-mnist: a novel image dataset for
  benchmarking machine learning algorithms,'' \emph{arXiv preprint
  arXiv:1708.07747}, 2017.

\bibitem{netzer2011reading}
\BIBentryALTinterwordspacing
Y.~Netzer, T.~Wang, A.~Coates, A.~Bissacco, B.~Wu, and A.~Y. Ng, ``Reading
  digits in natural images with unsupervised feature learning,'' 2011.
  [Online]. Available:
  \url{http://ufldl.stanford.edu/housenumbers/nips2011\_housenumbers.pdf}
\BIBentrySTDinterwordspacing

\bibitem{krizhevsky2009learning}
A.~Krizhevsky, G.~Hinton \emph{et~al.}, ``Learning multiple layers of features
  from tiny images,'' 2009.

\bibitem{liu2015faceattributes}
Z.~Liu, P.~Luo, X.~Wang, and X.~Tang, ``Deep learning face attributes in the
  wild,'' in \emph{Proceedings of International Conference on Computer Vision
  (ICCV)}, December 2015, pp. 3730--3738.

\bibitem{speechcommandsv2}
\BIBentryALTinterwordspacing
P.~{Warden}, ``{Speech Commands: A Dataset for Limited-Vocabulary Speech
  Recognition},'' \emph{ArXiv e-prints}, Apr. 2018. [Online]. Available:
  \url{https://arxiv.org/abs/1804.03209}
\BIBentrySTDinterwordspacing

\bibitem{simonyan2014very}
K.~Simonyan and A.~Zisserman, ``Very deep convolutional networks for
  large-scale image recognition,'' \emph{arXiv preprint arXiv:1409.1556}, 2014.

\bibitem{liu2020reactnet}
Z.~Liu, Z.~Shen, M.~Savvides, and K.-T. Cheng, ``Reactnet: Towards precise
  binary neural network with generalized activation functions,'' in
  \emph{European Conference on Computer Vision}.\hskip 1em plus 0.5em minus
  0.4em\relax Springer, 2020, pp. 143--159.

\bibitem{zheng2020learning}
R.~Zheng, Z.~Yu, Y.~Zhang, C.~Ding, H.~V. Cheng, and L.~Liu, ``Learning class
  unique features in fine-grained visual classification,'' \emph{arXiv preprint
  arXiv:2011.10951}, 2020.

\bibitem{cortes2012l2}
C.~Cortes, M.~Mohri, and A.~Rostamizadeh, ``L2 regularization for learning
  kernels,'' \emph{arXiv preprint arXiv:1205.2653}, 2012.

\bibitem{loshchilov2016sgdr}
I.~Loshchilov and F.~Hutter, ``Sgdr: Stochastic gradient descent with warm
  restarts,'' \emph{arXiv preprint arXiv:1608.03983}, 2016.

\bibitem{cubuk2019autoaugment}
E.~D. Cubuk, B.~Zoph, D.~Mane, V.~Vasudevan, and Q.~V. Le, ``Autoaugment:
  Learning augmentation strategies from data,'' in \emph{Proceedings of the
  IEEE/CVF Conference on Computer Vision and Pattern Recognition}, 2019, pp.
  113--123.

\bibitem{bulat2020bats}
\BIBentryALTinterwordspacing
A.~Bulat, B.~Martinez, and G.~Tzimiropoulos, ``Bats: Binary architecture
  search,'' in \emph{European Conference on Computer Vision (ECCV)}, 2020.
  [Online]. Available: \url{https://arxiv.org/abs/2003.01711}
\BIBentrySTDinterwordspacing

\bibitem{larq}
\BIBentryALTinterwordspacing
L.~Geiger and P.~Team, ``Larq: An open-source library for training binarized
  neural networks,'' \emph{Journal of Open Source Software}, vol.~5, no.~45, p.
  1746, Jan. 2020. [Online]. Available:
  \url{https://doi.org/10.21105/joss.01746}
\BIBentrySTDinterwordspacing

\bibitem{parmar2018image}
N.~Parmar, A.~Vaswani, J.~Uszkoreit, L.~Kaiser, N.~Shazeer, A.~Ku, and D.~Tran,
  ``Image transformer,'' in \emph{International Conference on Machine
  Learning}.\hskip 1em plus 0.5em minus 0.4em\relax PMLR, 2018, pp. 4055--4064.

\bibitem{zhang2019self}
H.~Zhang, I.~Goodfellow, D.~Metaxas, and A.~Odena, ``Self-attention generative
  adversarial networks,'' in \emph{International Conference on Machine
  Learning}.\hskip 1em plus 0.5em minus 0.4em\relax PMLR, 2019, pp. 7354--7363.

\bibitem{gao2020physical}
Y.~Gao, S.~F. Al-Sarawi, and D.~Abbott, ``Physical unclonable functions,''
  \emph{Nature Electronics}, vol.~3, no.~2, pp. 81--91, 2020.

\bibitem{gu2019flip}
C.~Gu, W.~Liu, Y.~Cui, N.~Hanley, M.~O'Neill, and F.~Lombardi, ``A flip-flop
  based arbiter physical unclonable function ({APUF}) design with high entropy
  and uniqueness for {FPGA} implementation,'' \emph{IEEE Transactions on
  Emerging Topics in Computing, DOI: 10.1109/TETC.2019.2935465}, 2019.

\bibitem{gao2018lightweight}
Y.~Gao, Y.~Su, L.~Xu, and D.~C. Ranasinghe, ``Lightweight (reverse) fuzzy
  extractor with multiple reference {PUF} responses,'' \emph{IEEE Transactions
  on Information Forensics and Security}, vol.~14, no.~7, pp. 1887--1901, 2018.

\bibitem{delange2021continual}
M.~Delange, R.~Aljundi, M.~Masana, S.~Parisot, X.~Jia, A.~Leonardis,
  G.~Slabaugh, and T.~Tuytelaars, ``A continual learning survey: Defying
  forgetting in classification tasks,'' \emph{IEEE Transactions on Pattern
  Analysis and Machine Intelligence}, 2021.

\bibitem{ramasesh2020anatomy}
V.~V. Ramasesh, E.~Dyer, and M.~Raghu, ``Anatomy of catastrophic forgetting:
  Hidden representations and task semantics,'' in \emph{International
  Conference on Learning Representations}, 2020.

\bibitem{gao2021noisfre}
Y.~Gao, Y.~Su, S.~Nepal, and D.~C. Ranasinghe, ``Noisfre: Noise-tolerant memory
  fingerprints from commodity devices for security functions,'' \emph{arXiv
  preprint arXiv:2109.02942}, 2021.

\end{thebibliography}

\end{document}